\documentclass[journal]{IEEEtran}

\usepackage{stfloats}
\usepackage{hhline}
\usepackage[utf8]{inputenc}
\usepackage{siunitx}
\usepackage{graphicx}
\usepackage{makecell}
\usepackage{comment}
\usepackage[dvipsnames]{xcolor}
\usepackage{adjustbox}
\usepackage{hyperref}
\usepackage{subcaption}
\usepackage{caption}
\usepackage{array}
\usepackage{algorithm,algorithmic}
\usepackage{multirow}
\usepackage{amsmath, amssymb}
\usepackage{color, colortbl}
\usepackage{booktabs}
\usepackage{amssymb}% http://ctan.org/pkg/amssymb
\usepackage{pifont}% http://ctan.org/pkg/pifont
\usepackage{soul}

\newcommand{\eg}{\textit{e.g., }}

\def \1{\textit{(i)}}
\def \2{\textit{(ii)}}
\def \3{\textit{(iii)}}
\def \4{\textit{(iv)}}
\def \5{\textit{(v)}}

\definecolor{beaublue}{rgb}{0.74, 0.83, 0.9}

\sethlcolor{yellow!50}

\begin{document}
\title{CyberForce: A Federated Reinforcement Learning Framework for Malware Mitigation}

\author{Chao Feng$^{*1}$, Alberto Huertas Celdr\'an$^{1}$, Pedro Miguel S\'anchez S\'anchez$^{2}$, Jan Kreischer$^{1}$, \\ Jan von der Assen$^{1}$, G\'er\^ome Bovet$^{3}$, Gregorio Mart\'inez P\'erez$^{2}$, and Burkhard Stiller$^{1}$

\thanks{$^{*}$Corresponding author.}

\thanks{This work has been partially supported by the Swiss Federal Office for Defense Procurement (armasuisse) with the DEFENDIS and CyberForce projects, and the University of Zürich UZH.}%
\thanks{$^{1}$Chao Feng, Alberto Huertas Celdr\'an, Jan Kreischer, Jan von der Assen, and Burkhard Stiller are with the Communication Systems Group (CSG) at the Department of Informatics (IFI), University of Zurich (UZH), 8050 Zürich, Switzerland {\tt\small (e-mail:[cfeng, huertas, vonderassen, stiller]@ifi.uzh.ch; jan.kreischer@uzh.ch}).}
\thanks{$^{2}$Pedro Miguel S\'anchez S\'anchez and Gregorio Mart\'inez P\'erez are with the Department of Information and Communications Engineering, University of Murcia, 30100 Murcia, Spain {\tt\small ([pedromiguel.sanchez, gregorio]@um.es)}.}%
\thanks{$^{3}$G\'{e}r\^{o}me Bovet is with the Cyber-Defence Campus within armasuisse Science \& Technology, 3602 Thun, Switzerland {\tt\small (gerome.bovet@armasuisse.ch)}.}}


\markboth{IEEE Transactions on Dependable and Secure Computing}%
{Shell \MakeLowercase{\textit{et al.}}: Bare Demo of IEEEtran.cls for IEEE Journals}

\maketitle

\begin{abstract} 
Recent research has shown that the integration of Reinforcement Learning (RL) with Moving Target Defense (MTD) can enhance cybersecurity in Internet-of-Things (IoT) devices. Nevertheless, the practicality of existing work is hindered by data privacy concerns associated with centralized data processing in RL, and the unsatisfactory time needed to learn right MTD techniques that are effective against a rising number of heterogeneous zero-day attacks. Thus, this work presents CyberForce, a framework that combines Federated and Reinforcement Learning (FRL) to collaboratively and privately learn suitable MTD techniques for mitigating zero-day attacks. CyberForce integrates device fingerprinting and anomaly detection to reward or penalize MTD mechanisms chosen by an FRL-based agent. The framework has been deployed and evaluated in a scenario consisting of ten physical devices of a real IoT platform affected by heterogeneous malware samples. A pool of experiments has demonstrated that CyberForce learns the MTD technique mitigating each attack faster than existing RL-based centralized approaches. In addition, when various devices are exposed to different attacks, CyberForce benefits from knowledge transfer, leading to enhanced performance and reduced learning time in comparison to recent works. Finally, different aggregation algorithms used during the agent learning process provide CyberForce with notable robustness to malicious attacks.

\end{abstract}


\IEEEpeerreviewmaketitle

\section{Introduction}

%IoT 

\IEEEPARstart{T}{he} rapid expansion of wireless communication technologies and the emergence of the Internet-of-Things (IoT) paradigm are leading to a substantial rise in the quantity of internet-connected devices with restricted capabilities. Currently, there are approximately 14 billion IoT devices, with projections indicating a rise to 64 billion by 2025~\cite{riad:2020:dynamic}, from healthcare to smart homes scenarios. These devices enhance human life and optimize productivity while minimizing costs.

%IoT and Cybersecurity isses
Although utilizing IoT devices brings numerous advantages, it also introduces specific cybersecurity concerns attributable to well-known and novel vulnerabilities found in resource-constrained devices \cite{meneghello:2019:threats}. One of the main concerns pertains to data privacy, as IoT devices typically gather sensitive information from users, such as location data, personal health records, or sensor readings. Another issue revolves around data integrity, as IoT devices rely on data sourced from various devices and users, ensuring the integrity and authenticity of the collected data is essential. Moreover, IoT devices are also vulnerable to various forms of attacks, including device compromise, malware injection, and network attacks. In such circumstances, malware can be distributed through compromised devices or malicious applications, potentially leading to data theft, unauthorized access, or control over the device \cite{sanchez2020survey}.

% MTD
Addressing all these aspects requires cybersecurity approaches that encompass secure communication protocols, data encryption, authentication mechanisms, intrusion detection systems, and continuous monitoring. Unfortunately, the complexity and resources consumed by these mechanisms, combined with zero-day vulnerabilities and attacks affecting resource-constrained devices, make them unsuitable for IoT platforms. Therefore, assuming the impracticality of perfect security, a pioneering cybersecurity paradigm known as Moving Target Defense (MTD) was introduced in 2009~\cite{cai2016moving}. MTD aims to counteract adversaries by proactively or reactively altering specific system parameters such as IP addresses, file extensions, or system libraries to impede and safeguard against attacks~\cite{JanVonDerAssen2022}.

% MTD and RL challenges
Proactive and reactive MTD approaches present important challenges that must be tackled. This work focuses on reactive approaches, where reinforcement learning (RL) has demonstrated its efficacy in learning which MTD mechanisms are able to mitigate heterogeneous zero-day attacks~\cite{huertas2022rl}. More in detail, for each device and attack, an RL-based agent acquires knowledge through trial and error, determining the most effective MTD mechanism based on discrepancies in the IoT device behavior before and after deploying each MTD mechanism. However, existing solutions require a remarkable amount of time to learn the right MTD per attack, and it is not scalable when several devices are under attack, making them unfeasible for many real IoT scenarios. The scalability issue is becoming even worse with the increasing number of zero-day attacks being faced by resource-constrained devices. In addition, existing RL-based solutions follow the traditional Machine Learning (ML) pipeline, where data is pooled in a central server, thus bringing concerns about data privacy when dealing with collaborative learning. These limitations could be tackled by Federated Learning (FL), where devices could collaborate to learn the MTD mechanisms mitigating each attack in a privacy-preserving manner. However, there is no work or solution that effectively integrates Federated and Reinforcement learning for the purpose of countering zero-day attacks with the utilization of MTD techniques. Therefore, critical performance and scalability aspects such as MTD selection accuracy, learning time, transfer of knowledge among IoT devices, and robustness against malicious attacks have not been compared to centralized RL-based approaches.

To address the previous challenges, the main contributions of the present work are:
\begin{itemize}
      \item {This work presents the design and implementation of CyberForce, an innovative cybersecurity framework that merges Federated and Reinforcement Learning to collaboratively learn optimal MTD techniques for diverse zero-day attacks while preserving privacy. It utilizes behavioral fingerprinting and ML-based anomaly detection to guide the decisions of an FRL-based agent, which employs Deep-Q Learning to master effective MTD techniques for various attack behaviors. Resource-limited devices can simultaneously train their models and share only parameters with the server, ensuring data privacy and efficient training through knowledge exchange. The framework's source code is publicly available {\cite{fedrlcode}}.}

    \item {This work details the deployment of CyberForce on resource-constrained Raspberry Pi 4 spectrum sensors within the real-world IoT platform Electrosense {\cite{electrosense}}. Each device faced six different malware attacks from ransomware, C\&C, and rootkit families. Four existing MTD mechanisms were utilized to combat these threats. This work develops two categories of real-world scenarios to investigate the influence of adversarial setups and data distributions on defensive strategies, which has yet to be previously explored in the literature. The two categories of scenarios include instances in which the attacker employs identical adversarial conditions across various clients, thereby representing an Independent and Identically Distributed (IID) data pattern, as well as instances where the attacker implements varying adversarial settings across different clients, reflecting a non-IID data distribution.}
    
    \item {Extensive experiments were conducted to assess CyberForce's effectiveness and learning time in mitigating the impact of malware attacks on the Electrosense sensor federation. These experiments encompassed various data distribution patterns, from IID to non-IID, from 10 clients to 20 clients, to thoroughly test CyberForce's adaptability and efficiency across diverse cybersecurity scenarios. The experimental findings indicated that the approach proposed in this paper effectively mitigated previously unseen malware by not only reducing the training time for the RL agent but also leveraging the knowledge-sharing capabilities inherent in FL.}

    \item {A pool of experiments evaluated CyberForce's resilience to poisoning attacks on the FRL-based agent. CyberForce employed different aggregation algorithms to mitigate these attacks. Experiments demonstrated that by adopting different aggregation functions, the framework proposed in this paper was able to resist many types of poisoning attacks. This paper discussed the trade-off between robustness and the model's performance. The Krum aggregation improved model robustness when facing model poisoning attacks in an IID environment, while FedAvg optimized performance in a secure environment with non-IID data.}  
\end{itemize}

The remainder of this article is structured as follows. Section \ref{sec:related} gives an overview of approaches that leverage RL, MTD, and FL. {The threat model is analyzed in Section {\ref{sec:threatmodel}}.} The design of CyberForce is introduced in Section~\ref{sec:framework}, and the experiments using its implementation are presented in Section~\ref{sec:experiments}. Finally, Section \ref{sec:conclusion} gives an overview of the conclusions and future research directions.

\section{Related Work}
\label{sec:related}

% RL, FL, or just MTD
This section reviews existing work focused on providing cybersecurity against a plethora of threats. \tablename~\ref{table:rw} shows and compares how RL, MTD, and FL have been considered by recent research to improve the security of different devices. 

In the IoT domain, \cite{reyFL2022} employed FL for anomaly detection and conducted experiments to demonstrate its superiority over traditional centralized ML methods, primarily due to its incorporation of data privacy protection. However, this solution did not consider malware mitigation and the usage of RL, as the work at hand does. {{\cite{nobakht2024sim}} introduced a FL-based malware detection model known as SIM-FED. This model addresses the challenges associated with data privacy and the constraints of limited resources in IoT devices. Furthermore, the research assessed the robustness of the SIM-FED model by conducting experiments involving poisoning attacks. {\cite{shukla2023federated}} proposed an FL approach designed to facilitate prompt updates to malware detection models. To tackle the issue of model heterogeneity prevalent in IoT systems, the proposed methodology employs a technique that translates heterogeneous models into a unified model within a cloud environment. {\cite{serpanos2023federated}} concentrated on the application of FL to facilitate the training of deep neural networks (DNNs) across various security operations centers (SOCs) belonging to different organizations to mitigate malware in a decentralized manner.} 

Another approach that leverages FL is~\cite{HyunKyoLim2020}, which advises how to combine FL with RL to collaboratively learn an optimal control policy. The collaboration amongst devices proved to accelerate the learning process, mitigate training instability and increase generalization. The main different with the paper at hand is that in the previous work, threats and MTD techniques were not considered. Similar to the work proposed in~\cite{HyunKyoLim2020}, ~\cite{AashmaUprety2021} provided evidence of optimizing defense with RL, however, the MTD paradigm is not supported. {{\cite{martinez2024mitigating}} implemented MTD strategies to enhance the security of edge computing systems. By utilizing techniques such as model parameter perturbation, topology shuffling, and algorithm switching, MTD effectively complicates the endeavors of potential attackers to exploit vulnerabilities within the system.}

\newcommand{\xmark}{\ding{55}}%
\setlength{\tabcolsep}{1.5pt}
\begin{table}[t]
\centering
\caption{Defense Approaches Leveraging RL, MTD, or FL}
\begin{tabular}{@{}llllccccc@{}}
\toprule
            \textit{Solution}         & \textit{Scenario}  & \textit{Device} &  \textit{Threat}&   \textit{Env.} & \textit{RL} & \textit{MTD} & \textit{FL}  & \textit{RA}\\ \midrule
    \cite{reyFL2022}  2022 & Network Security & IoT & Botnets & R & \xmark & \xmark & \checkmark & \xmark \\

    \cite{nobakht2024sim} 2024 & System Security & IoT & Malware & H & \xmark & \xmark & \checkmark & \checkmark\\
    
    \cite{shukla2023federated} 2023 & System Security & IoT & Malware & H & \xmark & \xmark & \checkmark & \xmark\\
    
    \cite{serpanos2023federated} 2023 & System Security & IoT & Malware & S & \xmark & \xmark & \checkmark & \xmark\\

    \cite{martinez2024mitigating} 2024 & Network Security & Edge & DDoS & H & \xmark & \checkmark & \checkmark & \xmark\\
    
    \cite{HyunKyoLim2020} 2020 & Optimal Control & IoT & None & R & \checkmark & \xmark & \checkmark & \xmark\\
    
    \cite{AashmaUprety2021} 2021 &  IT Security & IoT & DDoS, Spoof & S & \checkmark & \xmark & \xmark & \xmark\\

    \cite{KunSun2013}  2013 & Network Security & Servers & DDoS & S & \xmark & \checkmark & \xmark & \xmark\\

    \cite{TahaEghtesad2019} 2019 & Policy Planning & Servers & Probing & S & \checkmark & \checkmark & \xmark & \xmark\\
    
    \cite{SailikSengupta2020} 2020 & Web Security & Servers & Various & S & \checkmark & \checkmark & \xmark & \xmark\\

    \cite{KimSunghwan2021} 2021 & Intrusion Prevention & Network & DoS,  Scan & H & \checkmark & \checkmark & \xmark & \xmark\\
    
    \cite{MassimilianoAlbanese2018} 2018 & IT Security & Network & DDoS, Botnet & R & \checkmark & \checkmark & \xmark & \xmark\\

    \cite{XiaoyuXu2022} 2022 & Routing & SDN & Eavesdropping & R & \checkmark & \checkmark & \xmark & \xmark\\
    
    \cite{HengerLi2022} 2023 & System Security & CPS & From NVD & R & \checkmark & \checkmark & \xmark & \xmark\\

    \cite{ChungangGao2021} 2021 & Network Security & CPS & DDoS & S & \checkmark & \checkmark & \xmark & \xmark\\

    \cite{TaoZhang2023} 2023 & Network Security & IoV & DDoS & S & \checkmark & \checkmark & \xmark & \xmark\\
    
    \cite{SeunghyunYoon2021} 2021 & Network Security & IoV & Various & H & \checkmark & \checkmark & \xmark & \xmark\\

    \cite{huertas2022rl} 2022 & System Security & IoT & Malware & H & \checkmark & \checkmark & \xmark & \xmark\\

    \textit{This work} & Crowdsensing & IoT & Malware & R &\checkmark & \checkmark & \checkmark  & \checkmark \\
           \midrule
\end{tabular}%
\label{table:rw}
           Robustness Analysis (RA), Real-World (R), Simulated (S), Hybrid (H)
\end{table}

%\cite{KunSun2013} 

% RL + MTD
  % [in Servers]
Looking at the combination of RL and MTD, several approaches can be found in the literature that were implemented and validated in real or simulated environments. \cite{KunSun2013, TahaEghtesad2019, SailikSengupta2020} deployed MTD mechanisms on servers to mitigate various threats such as Distributed Denial-of-Service (DDoS) or Reconnaissance attacks. Although solutions such as \cite{TahaEghtesad2019} successfully demonstrated that RL can be used to find the optimal MTD technique, all three approaches were implemented in a simulated environment and none of them cover the applicability of FL to further optimize their approach, as the work at hand does. Moreover, the focus on computationally strong environments may indicate that they might not be suitable for resource-constrained devices.

  % [in the network]
With respect to network-based approaches, both generic elements of the attack surface (\eg IP addresses, TCP source ports) and Software-defined Networking (SDN) parameters are exploited to counter an array of threats, including Botnets, DDoS, and Reconnaissance attacks. \cite{KimSunghwan2021, MassimilianoAlbanese2018, XiaoyuXu2022} all combine RL and MTD to mitigate the threat vectors described. The validation of their techniques was conducted in real-world environments. However, none of them considered the effects of a federated setting with respect to their defense model. 

In the cyber-physical systems (CPS) domain, \cite{HengerLi2022} demonstrated that an MTD framework with RL can be used to pre-train policies by using simulated environments. More in detail, it was shown that MTD can be optimized to defend against unknown attacks. \cite{ChungangGao2021} proposed an RL-based mobile MTD strategy capable of balancing system security and system performance. The goal of the defender is to thwart DDoS attacks by launching a network shuffling MTD before the attacker completes the reconnaissance phase. Deep-Q learning was used to optimize and adapt to the evolving strategy of attackers. Experiments in a simulated environment have shown that this allows the defender to find a balance between security and performance. Similar approaches were followed by \cite{TaoZhang2023} and \cite{SeunghyunYoon2021}, using both simulated and real-world Internet-of-Vehicles (IoV) environments for evaluation. Although the above mentioned approaches demonstrate that RL-based MTD methods are effective in mitigating the impact of cyberattacks in multiple scenarios, none of them take into account the problem of data privacy. Finally, \cite{huertas2022rl} combined RL and existing MTD mechanisms to optimize the deployment of the techniques for zero-day attacks. An online RL agent was trained in a real-world scenario, resulting in a realistic validation scenario that was able to mitigate multiple samples from a wide range of malware families (\eg rootkits, ransomware, data extortion, or botnets). Concerning the learning environment, only an isolated device and agent were considered. Hence, no federated approaches were considered.

In summary, related approaches can be divided into three main categories. The first comprises works dealing with either MTD, RL, or FL in isolation. Only a few papers combine RL and FL to form the second category. Finally, numerous publications using RL for MTD deployment or optimization can be identified, constituting the third category. However, there is a gap in these studies regarding the consideration of data privacy. Additionally, {only a limited number of} existing work includes a thorough analysis of the robustness of their proposed approach. Thus, there is an opportunity to combine RL and FL for the deployment of MTD and to improve existing limitations. By exchanging the learned knowledge, devices that have not seen a specific attack can profit from behavioral learning. Furthermore, such a collaboration could reduce the overall training time, save resources for the devices, and preserve the privacy and security of the data.

% In summary, previous works have demonstrated the efficacy of integrating RL and MTD in mitigating the impact of malware attacks. However, there is a gap in these studies regarding the consideration of data security and privacy. Additionally, none of the existing work includes a thorough analysis of the robustness of their proposed approach. Strategies based on FL offer a promising solution to address both data privacy and security concerns. Therefore, there is an opportunity to integrate RL and FL in the deployment of MTD and improve upon the existing limitations. Moveover, FL offers a collaborative mechanism wherein clients are able to exchange their knowledge while gaining from the insights contributed by others. By exchanging the learned knowledge, devices that have not seen a specific attack can profit from behavioral learning. Thus, such a collaboration could reduce the overall training time, save resources for the devices, and preserve the data privacy and security.

%------------------------------------------------------------

\section{Threat Model}
\label{sec:threatmodel}
{A threat model is a structured process to identify and understand potential security threats or vulnerabilities in a system {\cite{threatfinderai}}. Therefore, it is necessary to perform threat modeling on the system before designing an FRL-based malware mitigation framework.

\textbf{Zero-day Attacks:} A zero-day attack is a cyberattack that takes advantage of an undiscovered vulnerability in software or hardware, resulting in the vendor having had "zero days" to develop a corrective patch or mitigation strategy {\cite{bilge2012before}}. The inherent risk of such attacks is heightened by the absence of immediate defensive measures. In this work, the device operates without prior knowledge regarding the attack, and the agent's selection of defensive measures is not pre-optimized for any particular threat, categorizing the attack as a zero-day attack. Furthermore, in a more realistic distributed environment, the attacks experienced by individual clients are different. Specifically, the attack faced by client \textit{i} lacks may not be faced by client \textit{j}, and no attack information is released from client \textit{i} to client \textit{j}. Thus, for client \textit{j}, this attack qualifies as a zero-day attack, as there is no prior knowledge or information about the attack for client \textit{j}. To address the challenge of zero-day attacks, this work designs and implements an FRL-based MTD mechanism. In this framework, the agent autonomously identifies the most effective defense strategy based on environmental feedback. To enhance applicability within distributed environments, this research facilitates knowledge sharing among clients in a privacy-preserving manner by integrating RL with the FL paradigm, thereby strengthening the mitigation of zero-day attacks.

\textbf{Attacker’s Objective:} This study examines three families of malware: C\&C systems, rootkits, and ransomware. These categories represent the most prevalent forms of malicious software and encompass a wide range of malicious activities {\cite{huertas2022rl}}. For C\&C attacks, attackers aim to gain unauthorized access, exfiltrate data, remotely control infected systems, move laterally within the network, and establish botnets for coordinated attacks. As for rootkit attacks, attackers focus on achieving stealth and persistence for long-term, undetected system access. They also seek privilege escalation, covert data exfiltration, and system manipulation to hide their presence further. For ransomware attacks, the primary objective is financial gain through encrypting valuable data and demanding a ransom payment in exchange for decryption. By leveraging the strengths of each attack type, they aim to maximize their financial gains while minimizing the risk of detection and recovery for the victim.

\textbf{Attacker’s Capabilities:} In these three malware families, attackers need technical proficiency in network protocols, operating systems, kernel-level programming, vulnerability exploitation, encryption, and evasion techniques. Social engineering skills may also be employed to trick users into installing malware. Specifically, in the attack considered in this work, the attacker has deployed the attack, by uncovering undisclosed vulnerabilities in devices in a distributed system, \textit{i.e.,} a zero-day attack. In the attacked devices, due to the lack of advanced disclosure of vulnerabilities from the vendors, it is impossible to predict the arrival of the attack in advance and take immediate response strategies.  

As previously mentioned, within a distributed environment, various clients encounter distinct threats; consequently, this study examines two distinct categories of attack scenarios. The first category posits that all nodes are susceptible to all types of attacks, meaning the attacker implements the aforementioned three families of attacks across all nodes, which share identical adversary settings. From a data perspective, the attack data matrix exhibits a similar distribution across clients, aligning with the concept of IID. In this scenario, RL can effectively automate the response strategy based on feedback, thereby mitigating the effects of the malware. Besides, the FL approach enhances the convergence rate of the model and facilitates the training of agents capable of addressing multiple attacks more efficiently. In the second attack scenario, attackers employ varying adversary settings tailored to different clients within the distributed system. In this scenario, the training dataset reveals that different clients encounter disparate attacks. From a data standpoint, this configuration results in a non-IID data distribution among individual clients. In such cases, FL offers a privacy-preserving mechanism for knowledge sharing by enabling the exchange of client models, thus equipping the agents with the ability to respond to a wide array of attack types. Thus, the following sections describe how to adopt an RL-based approach to dealing with zero-day attacks and how to combine the FL paradigm with a privacy-preserving approach through knowledge sharing to effectively defend against zero-day attacks in a distributed IoT environment.
}

\section{CyberForce Framework}
\label{sec:framework}

\begin{figure}[h]
\centering
\includegraphics[width=\linewidth]{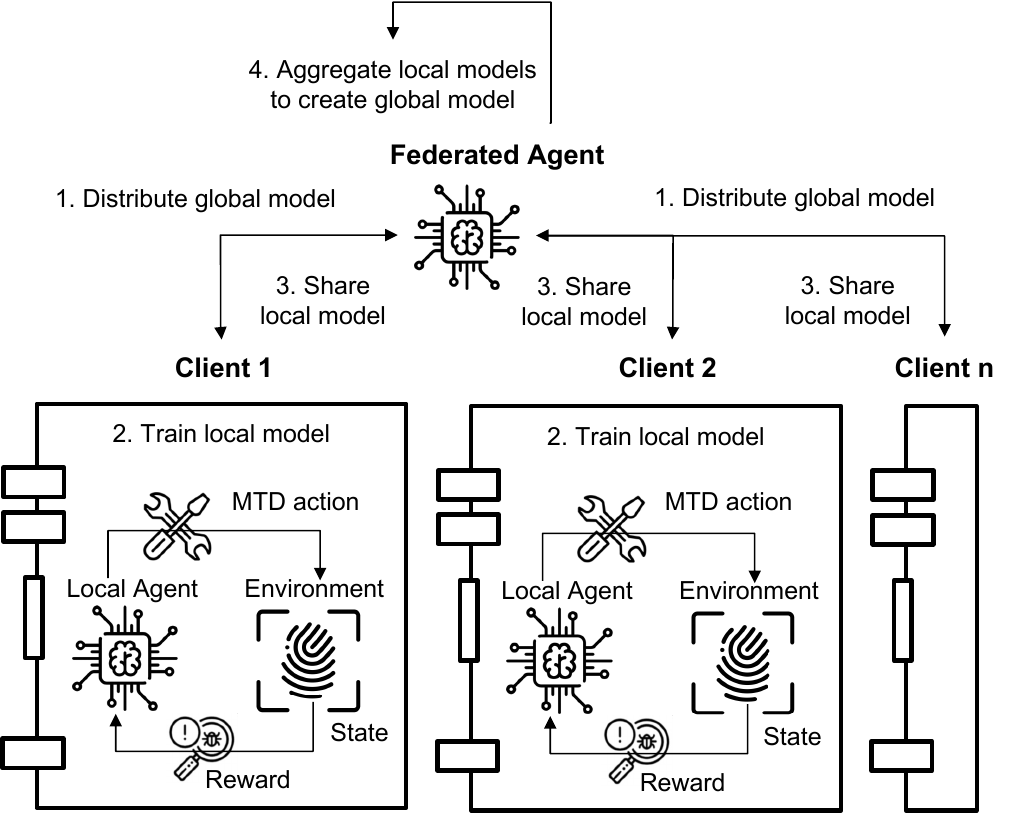}
\caption{CyberForce Framework Overview}
\label{fig:framework}
\end{figure}

This section introduces CyberForce, an FRL-based framework that learns optimal MTD mechanisms to mitigate unseen malware in a collaborative and privacy-preserving fashion~\cite{fedrlcode}. \figurename~\ref{fig:framework} shows the main elements and lifecycle of CyberForce. The main actor is the \textit{Federated Agent}, which combines FL and RL to learn effective MTD techniques mitigating heterogeneous zero-day attacks affecting various IoT devices. Each device of the federation hosts a \textit{Local Agent} that uses Deep Q-Learning to learn the right \textit{MTD} action according to the state of the device (called \textit{Environment}) and a \textit{Reward} mechanism that measures the impact of the MTD action on the environment.

\subsection{Agent and Federated Learning Process}

The FRL-based agent is the main novelty and contribution of this work. The agent learns MTD actions to mitigate attacks in an online fashion by using the Deep Q-learning algorithm and a federated neural network. First, the federated agent selects and distributes an initial neural network (see \figurename~\ref{fig:framework} and \tablename~\ref{tbl:autoencoder_hyperparameter} for hyperparameters configuration) among the devices of the federation. Then, in each client, the local agent trains its local neural network by interacting with the environment or client (step 2 in \figurename~\ref{fig:framework}). In this interaction, each local agent takes an MTD action for a given state. States are explained below and represent the agent's vision of the client at a given time. Then the new state (affected by the MTD action) is evaluated by a reward mechanism focused on anomaly detection (explained below), and the output is fed to the local agent. After that, the local agent selects the next MTD action based on this new information, and this loop is repeated. Sequences of the previous steps are called episodes. After a given number of episodes, the weights of the neural networks of each local agent are shared with the Federated Agent (step 3 in \figurename~\ref{fig:framework}), which aggregates them to create a global model (step 4 in \figurename~\ref{fig:framework}). CyberForce provides several algorithms, as shown in~\tablename~\ref{tab:aggregations}, that are used for local model aggregation, including FedAvg~\cite{mcmahan2017communication}, Krum~\cite{BrendanMcMahan2016}, and Trimmed Mean~\cite{yin2018byzantine}. After the aggregation, the global model is subsequently transmitted to each client, thereby replacing their respective local models. The previous steps are repeated for a given number of rounds until the federated neural network converges.

% Mathematically speaking, the goal of the FRL agent is to maximize the expected cumulative discounted future rewards for all clients: \begin{math} G_t=R_{t+1} + \gamma R_{t+2} + \gamma^2 R_{t+3} + ... = \sum_{k=0}^{\infty}R_{t+k+1} \end{math}. Where, \begin{math}R_{t}\end{math} denotes the reward at time step t, and \begin{math}\gamma\end{math} corresponds to a discount factor.
Mathematically, the goal of the FRL agent is to maximize the expected cumulative discounted future rewards for all clients, as shown in equation~\ref{eq:goal}.
\begin{center}
    \begin{equation}
        G_t=R_{t+1} + \gamma R_{t+2} + \gamma^2 R_{t+3} + ... = \sum_{k=0}^{\infty}R_{t+k+1}
        \label{eq:goal}
    \end{equation}
\end{center}
Where, \begin{math}R_{t}\end{math} denotes the reward at time step t, and \begin{math}\gamma\end{math} corresponds to a discount factor. 
To maximize this expected return \begin{math}G_t\end{math}, the federated agent needs to learn a policy. Deep FRL is a promising approach, as it approximates the action-value function via a deep neural network. Therefore, Federated Deep Q-Learning is an adequate choice as it utilizes Temporal Difference and accounts for large state spaces. 

\setlength{\tabcolsep}{2pt}

\begin{table}[h]
	\centering
	\caption{Aggregation Algorithms for FRL Framework}
        \begin{tabular}{l>{\raggedright\arraybackslash}m{7cm}}
        \toprule
            \textit{Aggregation} & \makecell[c]{\textit{Description}}  \\
            \hline
            FedAvg & It averages the parameters of all the local models to yield the aggregated global model, with using   
                        \begin{math}        
                        w = \sum_{k=1}^{K} \frac{1}{K} w_k
                        \setlength\abovedisplayskip{3pt}%shrink space
                        \setlength\belowdisplayskip{3pt}
                        \end{math}, where $w$ represents the weights of the updated global model, $w_k$ denotes the parameter of each local model, and $K$ signifies the number of clients.\\
            \cline{2-2}
            Krum &  It chooses the global model by identifying the local model with the highest similarity to the rest of the local models. The similarity is determined by calculating the inverse of Euclidean distance with: \begin{math}d =\sqrt{(w_{i}-w_{j})^{2}}\end{math}, where $w_i$ and $w_j$ denote the parameter of two local models.\\
            \cline{2-2}
            \begin{tabular}[c]{@{}l@{}}Trimmmed\\Mean\end{tabular} & It eliminates the outliers from the parameters of local models and subsequently calculates the average of the remaining values to get the global model.\\          
            \midrule
        \end{tabular}
        \label{tab:aggregations}
\end{table}

\subsection{Environment \& State}

% Environment
The environment consists of a set of devices affected by zero-day malware attacks. Particularly, this work considers ten Raspberry Pi 4 devices running an ElectroSense sensor~\cite{electrosense}. ElectroSense is a publicly accessible and open-source IoT crowdsensing platform that collects data on the radio frequency spectrum worldwide. In such a scenario, six types of malware, originating from the following three different families, have been identified as harmful for crowdsensing devices: C\&C, rootkits, and ransomware. \tablename~\ref{tab:malwares} summarizes the main aspects of the malware samples, more information can be found in \cite{huertas2022intelligent}.

\setlength{\tabcolsep}{2pt}

\begin{table}[h]
	\centering
	\caption{Behaviors of the Malware Affecting IoT Devices}
        \begin{tabular}{ll>{\raggedright\arraybackslash}m{5.5cm}}
        \toprule
            \textit{Malware} & \textit{Family} & \makecell[c]{\textit{Description}}  \\
            \hline
            The Tick & C\&C & It controls bots from a remote location through a server utilizing a remote shell and retrieving data from targeted devices \\
            \cline{3-3}
            Jakoritar & C\&C & It creates client and server components to facilitate the occurrence of data leakage and remote control \\
            % Backdoor\_jakoritar
            \cline{3-3}
            Dataleak & C\&C & It deploys a shell script, which utilizes the netcat command to regularly leak confidential data from either files or commands \\
            % Backdoor\_dataleak
            \cline{3-3}
            Beurk & rootkits & Its features range from hiding pseudo-terminal backdoor clients, files, directories, and real-time log cleanup to concealing processes, logins, and bypassing analysis \\
            \cline{3-3}
            Bdvl & rootkits & Its functionality is immense, and it ranges from hidden backdoors that allow multiple connection methods to keylogging and stealing passwords and files \\
            \cline{3-3}
            \begin{tabular}[c]{@{}l@{}}Ransom-\\ ware\_PoC\end{tabular}  & Ransomware & Crypto-ransomware with typical functionality, except that it is not controlled by a C\&C server \\
            \midrule
            % Ransomware\_PoC
        \end{tabular}
        \label{tab:malwares}
\end{table}

\begin{figure}[h]
\centering
\includegraphics[width=1\linewidth]{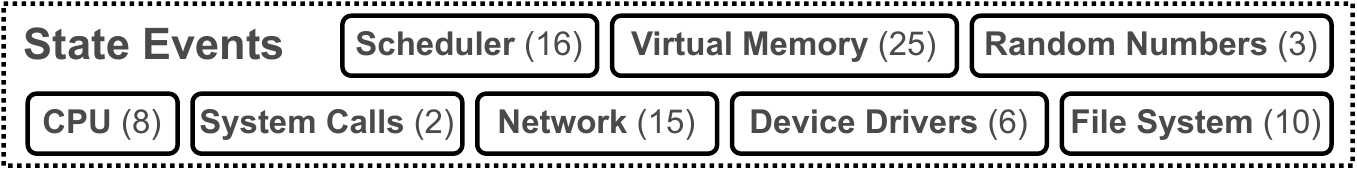}
\caption{CyberForce State Events}
\label{fig:state}
\end{figure}

% State
A state is the agent's vision of the environment at a given time. CyberForce uses device behavioral fingerprinting to represent environment states. In particular, software and kernel tracepoint events are considered because they cover an extensive range of promising events for attack detection and representation, as identified in previous works~\cite{huertas2022intelligent}. Initially, over 100 distinct \textit{perf} events were monitored, encompassing various dimensions such as system calls, CPU operations, device drivers, scheduler operations, network activities, file system operations, virtual memory usage, and random number generation. The selection criterion aimed to encompass a wide range of sources to identify minor disruptions caused by diverse zero-day attacks effectively. It is worth noting that the states or fingerprints should possess precise and stable characteristics over time, and the complexity of state representation (feature dimensionality) increases with a high number of events or features. Consequently, the learning process of the agent requires more time to converge. To address this, all features were monitored within time windows of 5 s over 8 hours (as suggested in \cite{huertas2022intelligent}), representing the normal behavior of Raspberry Pis. It is important to mention that previous studies have demonstrated the appropriateness of the chosen time window and monitoring duration \cite{huertas2022intelligent}. Once the dataset was collected, data distributions of all features were thoroughly examined. Features exhibiting constant or unstable values, as well as those with a correlation exceeding 90\%, were eliminated. Finally, a subset of 85 events was selected, as shown in \figurename~\ref{fig:state}.

\subsection{MTD Action}
The actions undertaken in this work pertain to the implementation of MTD techniques as a means to mitigate zero-day attacks. {Different malware has different targeted attack surface. When designing the MTD strategies that can be adopted by the agent, this work fully considers the targets of different families of malware. By dynamically adjusting the attack surface of the attacked device, it will increase the cost of the attacker to successfully implement the malware.} The primary objective of this framework is not to propose novel MTD mechanisms, but rather to establish a collaborative selection mechanism. Therefore, the framework considers the MTD techniques outlined in~\cite{JanVonDerAssen2022}. As shown in~\tablename~\ref{tbl:mtd_tech}, {in the context of C\&C attacks, which primarily target network infrastructures, this work utilizes a \textit{IP shuffling} strategy involving the dynamic alteration of IP addresses to mitigate the risk of exploitation of network vulnerabilities by adversaries.} {For ransomware, which predominantly focuses on the file system as its attack surface, two moving strategies, \textit{i.e.}, \textit{Ransomware trap} and \textit{File randomization} are implemented to safeguard the integrity of the file system.} \textit{Ransomware trap} creates dummy files that are subsequently encrypted by ransomware attacks. \textit{File randomization} modifies the file format extension, concealing the files from manipulation. {In the case of rootkits, where vulnerabilities are often exploited through system libraries, this work incorporates \textit{Library sanitation} to mitigate the effectiveness of rootkits.}  \textit{Library sanitation} shuffles shared system libraries between different sets and cleans associated links.

\begin{table}[h]
\centering
\caption{MTD Techniques for Malware Mitigation}
\label{tbl:mtd_tech}
\resizebox{\columnwidth}{!}{%
\begin{tabular}{lll}
\toprule
\textit{MTD Techniques}     & \textit{Mitigated Malware} & \textit{Malware Family} \\ \midrule
IP shuffling &  The Tick, Jakorita, Dataleak & C\&C \\
Ransomware trap & Ransomware\_PoC & Ransomware\\
File randomization & Ransomware\_PoC & Ransomware \\
Library sanitation & Beurk, Bdvl & rootkits\\ \midrule
\end{tabular}%
}
\end{table}

\subsection{Reward}

The learning process of the agent is facilitated by positive and negative rewards, which provide feedback on the efficacy of selected actions in different states. This study proposes the utilization of an anomaly detection (AD) system based on unsupervised ML to automate the reward generation process. Specifically, when an attack affects a client and a particular MTD technique is chosen by the local agent, the AD system evaluates the resulting device behavior. If the AD system predicts a normal behavior, it means that the deployed MTD technique effectively mitigated the attack, resulting in a positive reward. In contrast, if the device behavior is deemed abnormal, it indicates that the selected MTD technique was ineffective against the attack, leading to a negative reward.

To enable this functionality, an offline process is employed to train one Autoencoder per client using normal behavior. 

The training data is collected by monitoring the previously selected events over a period of eight days for each Raspberry Pi, which remained unaffected by any attacks. Subsequently, the datasets undergo several tasks, including: i) splitting into training and validation sets, ii) normalizing feature values, and iii) eliminating outliers using the Z-score method. Following this, individual Autoencoder models are trained for each device using 80\% of the dedicated normal data. The remaining 20\% of samples are utilized to calculate the threshold, determined by the mean predicted Mean Squared Error (MSE) reconstruction loss plus 2.5 standard deviations. Then, in real time, an online process evaluates each environment state. This involves executing malware samples on each Raspberry Pi and triggering the local agent if the AD system detects abnormal behavior. The local agent then selects and deploys a specific MTD technique. After giving the MTD technique two minutes to mitigate the attack, the AD system re-evaluates the device state. If the behavior is determined to be normal, a positive reward $(+1)$ is given to the agent. If the behavior remains abnormal, the reward is negative $(-1)$.

\section{Experiments}
\label{sec:experiments}

This section performs a pool of experiments to evaluate the learning process, mitigation performance, and robustness of CyberForce when different attacks affect various ElectroSense sensors and the federated learning process.

% More in detail, the first experiment evaluates the performance of the reward system, which influences the CyberForce learning time and  accuracy. Then, due to the federated nature of CyberForce, local agents are sensitive to different data distributions. Therefore, the second experiment evaluates three data distributions (from IID to non-IID). Finally, in FL, robustness is a critical criterion of trustworthiness. Thus, the last experiment analyzes the resiliency of CyberForce against malicious attacks, including data and model poisoning attacks.

\subsection{\textbf{Experiment 1}: Anomaly Detection for Rewards}

% Intrusion Detection System
The local RL agent is triggered when the AD determines that the current state of the device is abnormal. Therefore, the performance of the AD has a crucial influence on the overall effectiveness of the CyberForce framework. To identify zero-day attacks, it suffices to train the AD on normal behavioral patterns, enabling it to identify markedly different malware samples. Then, the agent is alerted and provided with the relevant state sample to choose an appropriate MTD.

\begin{table}[h]
\centering
\caption{Hyperparameter Search for the AD System}
\label{tbl:autoencoder_hyperparameter}
\resizebox{\columnwidth}{!}{%
\begin{tabular}{@{}lll@{}}
\toprule
\textit{\begin{tabular}[c]{@{}l@{}}Hyperparameter\\ Class\end{tabular}}                          & \textit{Type}        & \textit{Candidates}         \\ \midrule
\multirow{3}{*}{\begin{tabular}[c]{@{}l@{}}Model \\ Hyperparameter\end{tabular}}        
 & NR\_NEURONS\_PER\_LAYER & \textbf{(64, 32)}, (64, 16), (64, 8) \\
 & ACTIVATION\_FUNCTION & Sigmoid, Tanh, ReLU, ELU, \textbf{GELU} \\
 & BATCH\_NORMALIZATION & False, \textbf{True}                    \\\hline
\multirow{6}{*}{\begin{tabular}[c]{@{}l@{}}Optimization \\ Hyperparameter\end{tabular}} 
 & LOSS\_FUNCTION        & MAE, MSE, \textbf{RMSE}              \\
 & OPTIMIZER           & SGD, \textbf{Adam}, RMSprop             \\
 & LR                  & 1e-3, \textbf{1e-4}, 1e-5               \\
 & L2\_REGULARIZATION   & 1e-1, \textbf{1e-2}, 1e-3, 1e-4         \\
 & EARLY\_STOPPING      & False, \textbf{True (patience=5)}       \\
 & BATCH\_SIZE          & \textbf{32}, 64                         \\\hline
\begin{tabular}[c]{@{}l@{}}Prediction \\ Hyperparameter\end{tabular}                             & N\_STD                & 1, \textbf{2}, 3                     \\ \midrule
\end{tabular}%
}
\end{table}

This experiment performs a hyperparameter search for the AutoEncoder used as an AD model. The complete set of hyperparameters can be seen in \tablename~\ref{tbl:autoencoder_hyperparameter}, where a GridSearch and five-fold cross-validation were performed to test all possible hyperparameter combinations.
Additionally, early stopping with a patience of five is employed to prevent overfitting. The hyperparameter combination used by the best-performing AD model is displayed in bold in \tablename~\ref{tbl:autoencoder_hyperparameter}, which is used during the following experiments.

% \tablename~\ref{tbl:ae_best_hyperparameter_evaluation} shows the performance with the best hyperparameter combination of the AD for all states. Overall, this model achieves a 99.54\% accuracy for normal data and gets more than a 99\% successful detection rate for all kinds of malware. This result indicates that the AD system is able to effectively identify whether the device is under attack or not, and correctly trigger the agent to select the right strategy in time. The hyperparameter combination used by the best-performing AD model is displayed in bold in \tablename~\ref{tbl:autoencoder_hyperparameter}, which is used during the following experiments. The aim is to ascertain whether the presence of inaccurate rewards from the AD system affects the decision-making of agent or if the agent is unable to differentiate between normal states and zero-day attack states (or both scenarios).

\begin{table}[h]
\centering
\caption{AD Accuracy for States and Afterstates}
\label{tbl:ae_best_hyperparameter_evaluation}
\begin{tabular}{>{\raggedright\arraybackslash}m{5cm}>{\raggedright\arraybackslash}m{1.5cm}>{\raggedright\arraybackslash}m{1.5cm}}
\toprule
\textit{Behavior}                            & \textit{Accuracy} & \textit{Target State} \\
\hline
\textbf{Normal}                              & \textbf{99.54\%}  & \textbf{Normal}       \\
\hline
\textbf{Ransomware\_PoC (state)   }                  & \textbf{100.00\%} & \textbf{Abnormal}     \\
\textbf{Ransomware\_PoC + Ransomware trap }   & \textbf{100.00\%} & \textbf{Normal}       \\
\textbf{Ransomware\_PoC + File randomization} & \textbf{100.00\%} & \textbf{Normal}     \\
Ransomware\_PoC + IP shuffling       & 100.00\% & Abnormal     \\
Ransomware\_PoC + Library sanitation & 100.00\% & Abnormal     \\
\hline
\textbf{Bdvl (state) }                                & \textbf{99.52\%}  & \textbf{Abnormal}     \\
Bdvl + Ransomware trap              & 47.69\%  & Abnormal     \\
Bdvl + File randomization           & 48.29\%  & Abnormal     \\
Bdvl + IP shuffling                 & 59.68\%  & Abnormal     \\
\textbf{Bdvl + Library sanitation}           & \textbf{99.51\%}  & \textbf{Normal}       \\
\hline
\textbf{Beurk   (state)  }                            & \textbf{99.89\%}  & \textbf{Abnormal}     \\
Beurk + Ransomware trap             & 0.00\%   & Abnormal     \\
Beurk + File randomization          & 0.10\%   & Abnormal     \\
Beurk + IP shuffling                & 0.10\%   & Abnormal     \\
\textbf{Beurk + Library sanitation   }       & \textbf{100.00\%} & \textbf{Normal}       \\
\hline
\textbf{The Tick    (state)  }                        & \textbf{99.01\%}  & \textbf{Abnormal}     \\
The Tick + Ransomware trap          & 0.09\%   & Abnormal     \\
The Tick + File randomization       & 0.10\%   & Abnormal     \\
\textbf{The Tick + IP shuffling   }          & \textbf{99.61\%}  & \textbf{Normal}       \\
The Tick + Library sanitation       & 0.00\%   & Abnormal     \\
\hline
\textbf{Jakoritar    (state) }                        & \textbf{99.76\%}  & \textbf{Abnormal }    \\
Jakoritar + Ransomware trap         & 0.00\%   & Abnormal     \\
Jakoritar + File randomization      & 0.32\%   & Abnormal     \\
\textbf{Jakoritar + IP shuffling  }          & \textbf{100.00\%} & \textbf{Normal}       \\
Jakoritar + Library sanitation      & 0.00\%   & Abnormal     \\
\hline
\textbf{Dataleak   (state)}                          & \textbf{99.56\%}  & \textbf{Abnormal}     \\
Dataleak + Ransomware trap          & 0.00\%   & Abnormal     \\
Dataleak + File randomization       & 0.00\%   & Abnormal     \\
\textbf{Dataleak + IP shuffling}             & \textbf{99.51\%}  & \textbf{Normal}       \\
Dataleak + Library sanitation       & 0.00\%   & Abnormal    \\
\midrule
\end{tabular}%
\end{table}

% \begin{table}[h]
%     \centering
%     \caption{Accuracy Score of the Anomaly Detector}
%     \resizebox{\columnwidth}{!}{%
%      \begin{tabular}{llllllll}
%         \toprule
%          Normal & Bdvl & Beurk & Backdoor & The Tick & Dataleak  & \begin{tabular}[c]{@{}l@{}}Ransom-\\ ware\_PoC\end{tabular}  \\
%         \midrule
%          99.54\% & 99.52\% & 99.89\% & 99.76\% & 99.01\%  & 99.56\% & 99.28\% \\
%         \midrule
%     \end{tabular}
%     }
%     \label{tbl:ae_best_hyperparameter_evaluation}
% \end{table}

This experiment evaluates the performance of the AD system in two aspects: (i) on the behavior of the device which it is affected by attacks; and (ii) when subsequent deployed the MTD techniques, including both correct ones and incorrect ones. If the deployment of the appropriate MTD for each attack leads to the device returning to a normal state, it indicates that the AD system effectively provides precise feedback to the agent. Similarly, if the AD system detects abnormal behavior in cases where an incorrect MTD strategy is applied to an attack, this also implies that the AD is capable of providing accurate feedback to the agent. \tablename~\ref{tbl:ae_best_hyperparameter_evaluation} shows the performance with the best hyperparameter combination of the AD for all states. The results present the effectiveness of the AD in identifying anomalies when considering the current states alone, as well as when taking into account the subsequent outcomes of implementing MTD techniques. The first column displays the behavior of the device and the applied MTD technique. The accuracy of detecting normal or abnormal behavior is reflected in the second and third columns. All states that adhere to proper MTD techniques for a specific attack should be identified as normal, whereas state that employ incorrect or ineffective MTD strategies should be abnormal. These results are obtained evaluating approximately 1000 samples per behavior.

Overall, the model achieves a 99.54\% accuracy for normal behavior and gets more than a 99\% successful detection rate for all kinds of malware through the analysis of current device behavior. When it comes to accuracy of AD system with device behavior after the implementation of MTD techniques, two primary observations can be made. Firstly, the accuracy is comparable to that attained for states. However, if an incorrect MTD technique is implemented, the recognition of the state is notably poor in the case of Beurk, Dataleak, Jakoritar, and The Tick, with an accuracy of almost 0, while an acceptable accuracy score for Ransomware\_PoC and Bdvl. It implies  that the AD can offer precise feedback to the agent by detecting abnormal behavior when an incorrect MTD strategy is used against an attack. The second aspect is that, by implementing appropriate MTD techniques, all behaviors are correctly identified with a precision that surpasses 99\%. The AD system effectively provides precise feedback to the agent, as each attack is countered with the appropriate MTD, resulting in the device returning to its normal state. In conclusion, these results indicate that the AD system demonstrates its effectiveness in several crucial aspects. It can accurately identify whether the device is currently under attack, promptly trigger the agent to select the right strategy in time, and provide precise feedback regarding the actions decided by the agent.

\subsection{\textbf{Experiment 2}: FRL-based Agent for MTD Selection}

\subsubsection{Agent Configuration}
The Deep Q-Learning model holds utmost importance within the proposed CyberForce framework, as its ability to make accurate decisions based on the current device state is crucial. The performance of the model is greatly influenced by the hyperparameters, therefore, identifying the optimal combination of hyperparameters would prove advantageous for future experiments. This experiment lists 17 different hyperparameters from three aspects of the federation, neural network, and training strategy to find the most suitable combination, as shown in \tablename~\ref{tbl:deepqlearning_hyperparameter}.

\begin{table}[h]
\centering
\caption{Hyperparameter Search for FRL-based Agent}
\label{tbl:deepqlearning_hyperparameter}
\resizebox{\columnwidth}{!}{%
\begin{tabular}{@{}lll@{}}
\toprule
\textit{Element} & \textit{Hyperparameter}  & \textit{Values}         \\ \midrule

\multirow{5}{*}{\begin{tabular}[c]{@{}l@{}}Federation \end{tabular}}       
 % & NR\_CLIENTS & \textbf{10} \\
 & NR\_ROUNDS & \textbf{30} \\
 & NR\_EPISODES\_PER\_ROUND & \textbf{100} \\
 & NR\_EPISODES\_PER\_CLIENT & \textbf{3,000} \\
 & TOTAL\_NR\_EPISODES & \textbf{30,000} \\\hline

\multirow{3}{*}{Neural Network} 
 & NR\_NEURONS\_PER\_LAYER & \textbf{(128, 64)}, (128, 64, 32), (128, 64, 32, 16) \\
 & ACTIVATION\_FUNCTION & Sigmoid, Tanh, ReLU, \textbf{SELU} \\
 & DROPOUT                  & \textbf{0}, 0.2, 0.5               \\
\hline

\multirow{9}{*}{Training}                    & OPTIMIZER           & SGD, \textbf{Adam}, RMSprop, Adagrad           \\     
  & LOSS\_FUNCTION        & MAE, MSE, \textbf{RMSE}              \\   
  & GAMMA                & 0.1, 0.2, 0.3, 0.4, \textbf{0.5}, 0.6, 0.7, 0.8, 0.9  \\
  & LEARNING\_RATE               & 1e-2, 1e-3, \textbf{1e-4}, 1e-5  \\ 
  & L2\_REGULARIZATION               & 0, 1e-1, \textbf{1e-2}, 1e-3, 1e-4 \\ 
  & EPSILON\_START               & \textbf{1.0 } \\ 
  & EPSILON\_DEC               & \textbf{0.8/NR\_EPISODES\_PER\_CLIENT}  \\ 
  & EPSILON\_END                & \textbf{0.01}  \\ 
  & AGGREGATION\_STRATEGY                & \textbf{FedAvg}  \\ 
  \midrule
\end{tabular}%
}
\end{table}

\begin{table}[h]
\centering
\caption{Configuration of Experiment II}
\label{tab:exp2_config}
\resizebox{\columnwidth}{!}{%
\begin{tabular}{lll}
\toprule
\multicolumn{2}{l}{\textbf{CONFIGURATION}} & \textbf{VALUE} \\ \midrule
\multirow{5}{*}{\textbf{Federation}} & Number of Clients & 10, 20 \\ \cmidrule(l){2-3}
 & Total Rounds & 30 \\ \cmidrule(l){2-3}
 & Episodes in Each Round & 100 \\ \cmidrule(l){2-3}
 & Aggregation Function& \textit{FedAvg}, \textit{Krum}, \textit{Trimmed Mean} \\ \hline 
\multirow{3}{*}{\textbf{Attacks}} & Malware & \begin{tabular}[c]{@{}l@{}}The Tick, Jakoritar, Dataleak\\ Beurk,  Bdvl, Ransomware\_PoC\end{tabular} \\ \cmidrule(l){2-3}
 & Data Distribution & IID, non-IID (weak, strong) \\ \cmidrule(l){2-3}
  & Percentage of Malware Missing Clients & 0, 10\%, 40\%, 70\%, 100\% \\ \hline 
\end{tabular} 
}
\end{table}

Through the five-fold cross-validation strategy, this work tests all possible combinations. For this, 30,000 training samples are assigned to 10 clients and subsequently, each client is trained with 3,000 episodes distributed over 30 rounds of FL with 100 episodes per round. 

The best hyperparameter combinations are bolded in \tablename~\ref{tbl:deepqlearning_hyperparameter}. {For the local agent, a two-layer fully connected neural network architecture with dimensions of 128 * 64 is employed as the topology for the Deep Q-Learning framework. The activation function chosen for this model is the Scaled Exponential Linear Unit (SELU), and dropout is not applied. From a training perspective, the Adam optimization algorithm is utilized, and the Root Mean Square Error (RMSE) is adopted as the loss function. A learning rate of 1e-4 is established. To reduce the risk of overfitting, L2 regularization is implemented with a coefficient set to 1e-2. In the FL process, each participating agent conducts training of its local model over a predetermined duration of 100 episodes per round. Subsequently, these local models are transmitted to a central server for model aggregation, utilizing the aggregation function specified in TABLE {\ref{tab:aggregations}}. Following the aggregation process, the updated global model is sent back to the client side, thereby replacing the existing local models with the newly generated global model. This iterative process is repeated for a total of 30 rounds. These agent setups} are used for the following experiments.

\subsubsection{{Experimental Settings}}
{Experiment II aims to assess the effectiveness of the proposed FRL framework in addressing various families of malware across different adversarial settings, as listed in TABLE {\ref{tab:exp2_config}}. To this end, two categories of scenarios are examined to analyze the influence of adversarial settings and data distribution on the FRL framework. The first scenario centers on a uniform adversarial environment, wherein all nodes within the distributed system experience the same types of malware, implying that the training data adheres to an IID configuration. This scenario seeks to evaluate whether the FRL framework can deliver a defense comparable to that of traditional RL-based methodologies {\cite{huertas2022rl}}, as well as to determine whether FRL can confer additional advantages in terms of training efficiency, specifically by reducing the time necessary for the global model to achieve convergence.

In the second category of scenario, characterized by heterogeneous adversarial environments wherein nodes experience non-IID conditions, this experiment examines two distinct cases. In the first case, a predetermined number of clients are subjected to specific malware attacks. For instance, within a distributed system comprising ten nodes, clients 1 through 9 are targeted by the six types of malware enumerated in TABLE {\ref{tab:malwares}}, whereas client 10 is attacked by only five of these, excluding Ransomware\_PoC. This experimental setup aims to assess whether the FRL framework can effectively learn to mitigate these attacks without prior exposure to zero-day threats, leveraging knowledge transfer among clients. In the second non-IID case, each node is randomly subjected to three distinct attack types. This experimental design aims to evaluate the system's capacity to maintain effective defensive measures despite the significant variability in the adversarial conditions encountered by the nodes.
}

\subsubsection{IID Scenario}
\begin{figure}[H]
    \centering
    \subfloat[\centering CyberForce with 10 Clients]{{\includegraphics[height=2.9cm]{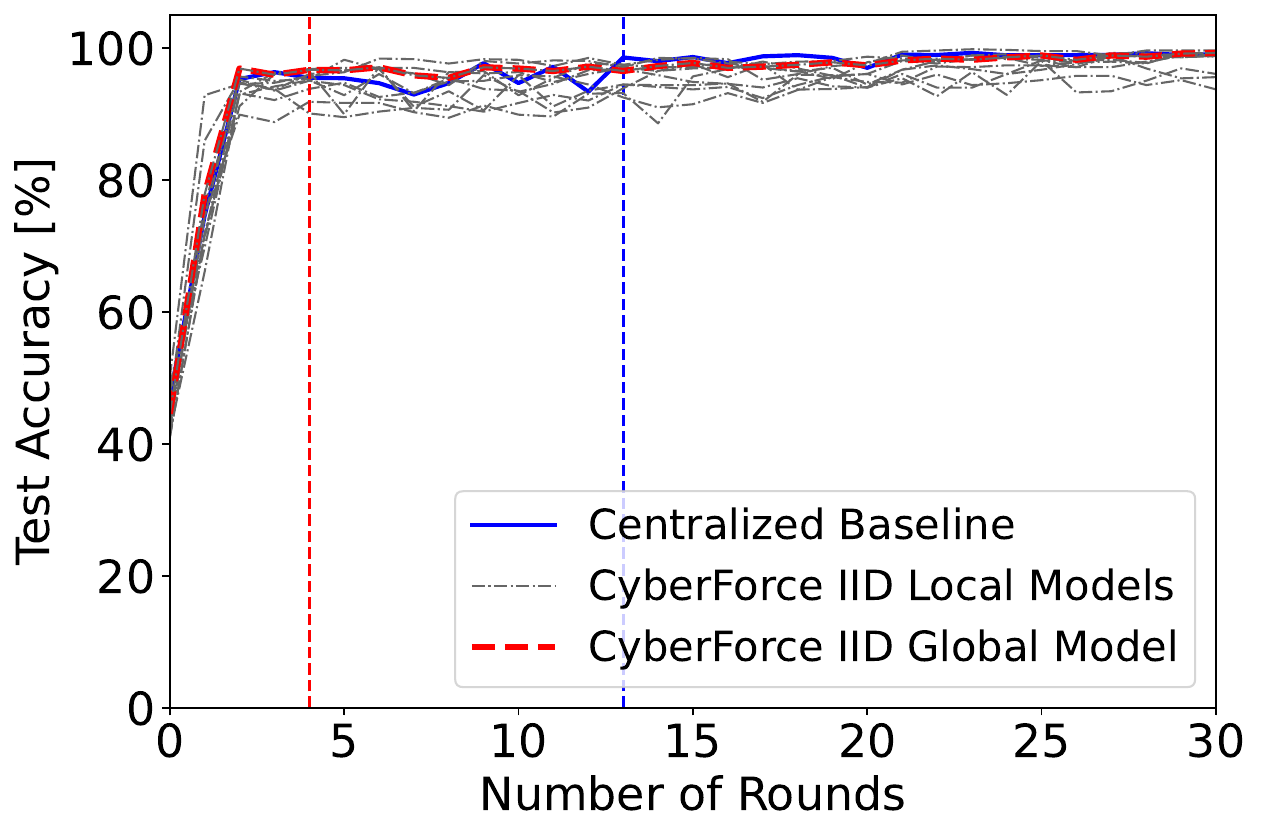} }}%
    \subfloat[\centering CyberForce with 20 Clients]{{\includegraphics[height=2.9cm]{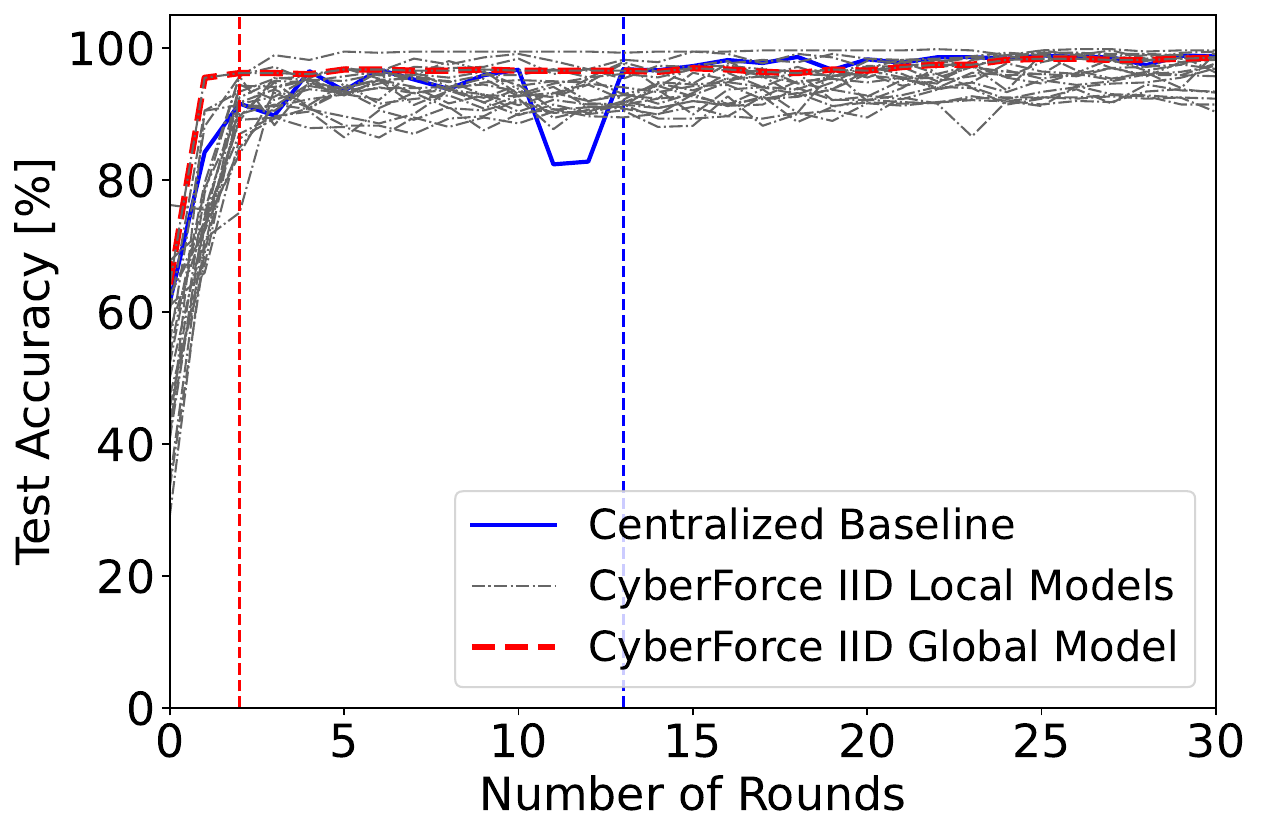} }} \\
\caption{Test Accuracy of Centralized Baseline~\cite{huertas2022rl} and CyberForce with IID Data Distribution}
\label{fig:baseline_learningcurve}
\end{figure}

To measure the effectiveness of the FRL training approach, the malware mitigation evaluation begins with the ideal scenario where all clients have the same and balanced data of each malware type (data follows an IID distribution across the clients). {In this investigation, the data was randomly averaged for each node, ensuring that each node was exposed to all forms of attacks. The experiment implemented two distinct configurations consisting of 10 and 20 nodes to assess the scalability of the CyberForce framework.} As a comparison, this work conducts a Centralized Baseline model according to~\cite{huertas2022rl}, where all the data is located in a single client and the RL model is trained based on a traditional ML pipeline. The accuracy score of Centralized Baseline and CyberForce with IID are shown in \figurename~\ref{fig:baseline_learningcurve}. As can be seen, CyberForce has a comparable malware mitigation efficiency with Centralized Baseline, both achieve more than 98\% accuracy at the end of the learning process, {no matter whether 10 or 20 clients are used}. {However, the federated global model of CyberForce demonstrates an accuracy of 96\% in round 4 with the utilization of 10 clients, as illustrated in Fig.{\ref{fig:baseline_learningcurve}}(a), which corresponds to a total of 400 episodes. Additionally, the global model achieves a similar accuracy of 96\% in round 2 when 20 clients (200 episodes) are engaged, depicted in Fig.{\ref{fig:baseline_learningcurve}}(b). As a comparison, }the Centralized Baseline achieves 96\% accuracy at 1300 episodes. Thus, it can be seen that CyberForce is able to save about two-thirds of the learning time compared to the traditional Centralized training approach, such as the one proposed in~\cite{huertas2022rl}.

\subsubsection{Non-IID Scenarios}
More realistically, data is not uniformly distributed across clients, not all clients are exposed to all malware, as it is essential to evaluate the performance of the framework in the non-IID case. This work designs two different non-IID setups: weak non-IID and strong non-IID.

In the weak non-IID setup, only one type of malware does not appear in the training dataset for one client at a time. For example, in a federation of ten clients, only client one has not been exposed to The Tick malware, while the remaining nine clients are exposed to it. To verify the impact of such a non-IID setup on the CyberForce framework, this experiment evaluates the accuracy of the absent malware. In other words, the mitigation accuracy for The Tick in the aforementioned example. Moreover, to validate the impact of the extent of malware-missing clients on the collaboration of CyberForce, the experiment sequentially increases the {ratio} of malware-missing clients from {10\%} to {100\%}. 

% \usepackage{multirow}
% \usepackage{graphicx}
\begin{table}[h]
\centering
\caption{Absent Malware Accuracy for Weak Non-IID (CyberForce with 10 Clients)}
\label{tbl:weak_non_iid}
\resizebox{\columnwidth}{!}{%
\begin{tabular}{llllll}
\midrule
                                &              & \multicolumn{4}{l}{\textit{No. of Malware Missing Clients}} \\
\hline                                
\textit{Absent Malware  }                & \textit{Algorithm}    & 1           & 4          & 7          & 10         \\
\hline
\multirow{3}{*}{The Tick}       & FedAvg       & 100.00\%    & 100.00\%   & 100.00\%   & 100.00\%   \\
                                & Krum         & 100.00\%    & 100.00\%   & 100.00\%   & 100.00\%   \\
                                & Trimmed Mean & 100.00\%    & 100.00\%   & 100.00\%   & 100.00\%   \\
                                \cline{2-6}
\multirow{3}{*}{Jakoritar}      & FedAvg       & 99.74\%     & 99.48\%    & 100.00\%   & 100.00\%   \\
                                & Krum         & 97.40\%     & 0.00\%     & 0.00\%     & 0.00\%     \\
                                & Trimmed Mean & 98.70\%     & 33.50\%    & 0.00\%     & 0.00\%     \\
                                \cline{2-6}
\multirow{3}{*}{Dataleak}       & FedAvg       & 95.68\%     & 92.15\%    & 89.21\%    & 94.11\%    \\
                                & Krum         & 100.0\%     & 14.37\%    & 6.53\%     & 10.45\%    \\
                                & Trimmed Mean & 99.01\%     & 98.03\%    & 80.39\%    & 9.60\%     \\
                                \cline{2-6}
\multirow{3}{*}{Beurk}          & FedAvg       & 96.15\%     & 95.05\%    & 96.93\%    & 94.21\%    \\
                                & Krum         & 96.71\%     & 0.00\%     & 0.00\%     & 0.00\%     \\
                                & Trimmed Mean & 95.60\%     & 4.24\%     & 0.00\%     & 0.00\%     \\
                                \cline{2-6}
\multirow{3}{*}{Bdvl}           & FedAvg       & 97.52\%     & 96.68\%    & 97.95\%    & 94.93\%    \\
                                & Krum         & 99.34\%     & 0.32\%     & 0.00\%     & 0.00\%     \\
                                & Trimmed Mean & 100.0\%     & 98.62\%    & 72.74\%    & 0.00\%     \\
                                \cline{2-6}
\multirow{3}{*}{Ransomware-PoC} & FedAvg       & 99.40\%     & 98.81\%    & 98.81\%    & 33.25\%    \\
                                & Krum         & 100.0\%     & 99.40\%    & 32.54\%    & 0.00\%     \\
                                & Trimmed Mean & 99.40\%     & 99.40\%    & 98.10\%    & 0.59\%    \\
                                \hline
\end{tabular}%
}
\end{table}
% \usepackage{multirow}
% \usepackage{graphicx}
\begin{table}[h]
\centering
\caption{Absent Malware Accuracy for Weak Non-IID (CyberForce with 20 Clients)}
\label{tbl:weak_non_iid_20clients}
\resizebox{\columnwidth}{!}{%
\begin{tabular}{llllll}
\midrule
                                &              & \multicolumn{4}{l}{\textit{No. of Malware Missing Clients}} \\
\hline                                
\textit{Absent Malware  }                & \textit{Algorithm}    & 2           & 8          & 14          & 20         \\
\hline
\multirow{3}{*}{The Tick}       & FedAvg       & 100.00\%	& 100.00\%	& 100.00\%	& 100.00\%   \\
                                & Krum         & 100.00\%	& 99.44\%	& 100.00\%	& 100.00\%   \\
                                & Trimmed Mean & 100.00\%	& 100.00\%	& 100.00\%	& 100.00\%   \\
                                \cline{2-6}
\multirow{3}{*}{Jakoritar}      & FedAvg       & 97.50\%	& 45.50\%	& 0.00\%	& 0.00\%   \\
                                & Krum         & 100.00\%	& 98.50\%	& 0.00\%	& 0.00\%     \\
                                & Trimmed Mean & 99.00\%	& 15.00\%	& 0.00\%	& 0.00\%     \\
                                \cline{2-6}
\multirow{3}{*}{Dataleak}       & FedAvg       & 100.00\%	& 100.00\%	& 95.90\%	& 0.00\%    \\
                                & Krum         & 100.00\%	& 72.82\%	& 0.00\%	& 0.00\%   \\
                                & Trimmed Mean & 100.00\%	& 100.00\%	& 86.15\%	& 0.00\%     \\
                                \cline{2-6}
\multirow{3}{*}{Beurk}          & FedAvg       & 88.41\%	& 21.16\%	& 0.00\%	& 0.00\%    \\
                                & Krum         & 98.84\%	& 99.71\%	& 0.00\%	& 0.00\%     \\
                                & Trimmed Mean & 97.10\%	& 15.94\%	& 1.45\%	& 0.00\%     \\
                                \cline{2-6}
\multirow{3}{*}{Bdvl}           & FedAvg       & 100.00\%	& 100.00\%	& 36.98\%	& 0.00\%   \\
                                & Krum         & 100.00\%	& 0.38\%	& 0.00\%	& 0.00\%     \\
                                & Trimmed Mean & 99.62\%	& 97.36\%	& 74.72\%	& 0.00\%     \\
                                \cline{2-6}
\multirow{3}{*}{Ransomware-PoC} & FedAvg       & 98.86\%	& 95.45\%	& 96.59\%	& 80.91\%   \\
                                & Krum         & 98.86\%	& 98.86\%	& 0.00\%	& 30.23\%     \\
                                & Trimmed Mean & 98.86\%	& 100.00\%	& 98.86\%	& 0.00\%   \\
                                \hline
\end{tabular}%
}
\end{table}

\tablename~\ref{tbl:weak_non_iid} presents the test accuracy of absent malware data for weak non-IID configured clients with three aggregation algorithms{, when the federation consists of ten clients}. In general, FedAvg demonstrates superior performance in addressing weak non-IID scenarios, specifically when the number of malware-missing clients is below ten. In such cases, FedAvg effectively learns all absent malware behavior. This can be attributed to FedAvg exceptional capacity to collectively acquire and distribute knowledge from all participating clients. Krum, however, is unable to effectively deal with the challenges posed by non-IID. When the number of malware missing clients increases, Krum has a higher chance of selecting a client that is not exposed to the absent malware, resulting in an overall federation that lacks data for that absent malware and cannot effectively respond to absent attacks. Additionally, Trimmed Mean shows a balanced performance. When the percentage of malware missing client is less than 40\%, Trimmed Mean is almost equivalent to FedAvg. However, when the percentage of malware missing client is greater than 70\%, the normal clients who contain all the malware behavior instead become outliers and are excluded by the Trimmed Mean, rendering the system incapable of effectively responding to absent malware.

\begin{figure}[t]
    \centering
    \subfloat[\centering 10\% of  Rootkit Missing Client]{{\includegraphics[height=3cm]{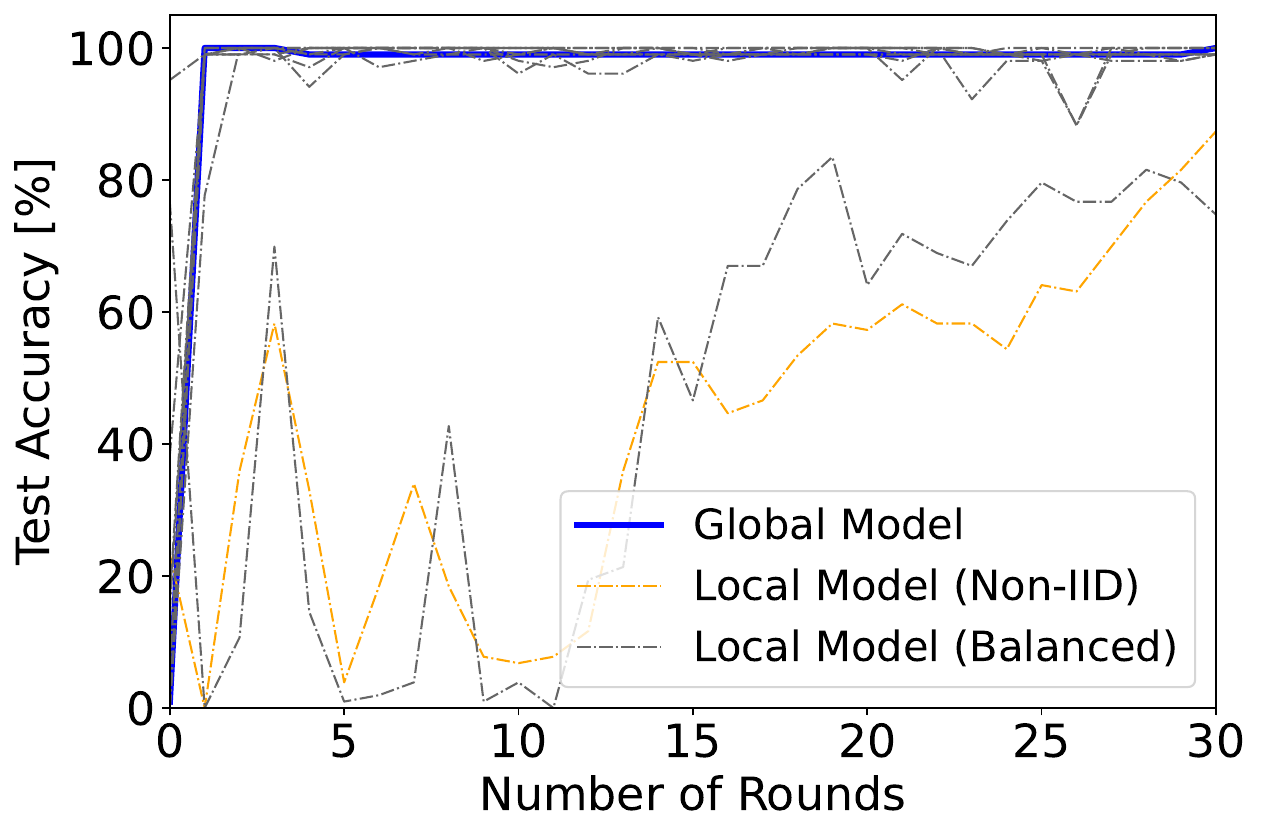} }}%
    \subfloat[\centering 70\% of  Rootkits Missing Client]{{\includegraphics[height=3cm]{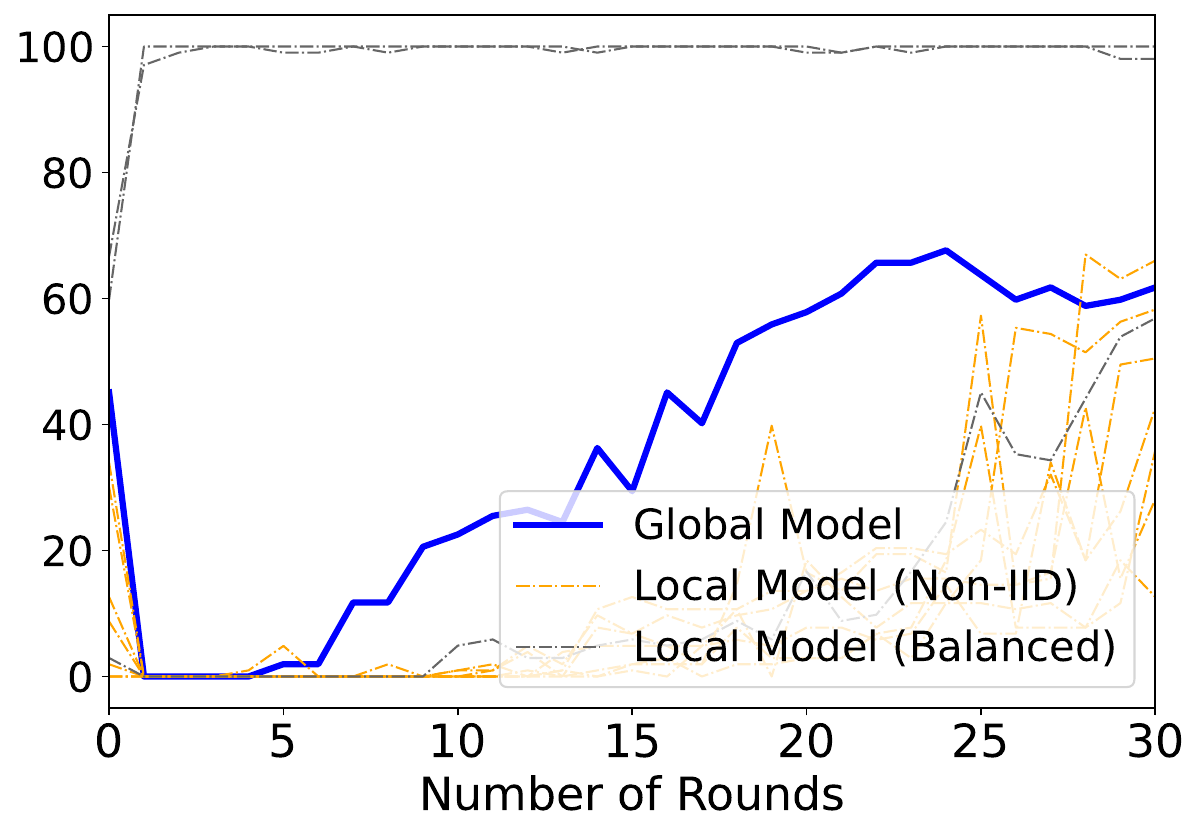} }} \\
    \subfloat[\centering 10\% of  C\&C Missing Client]{{\includegraphics[height=3cm]{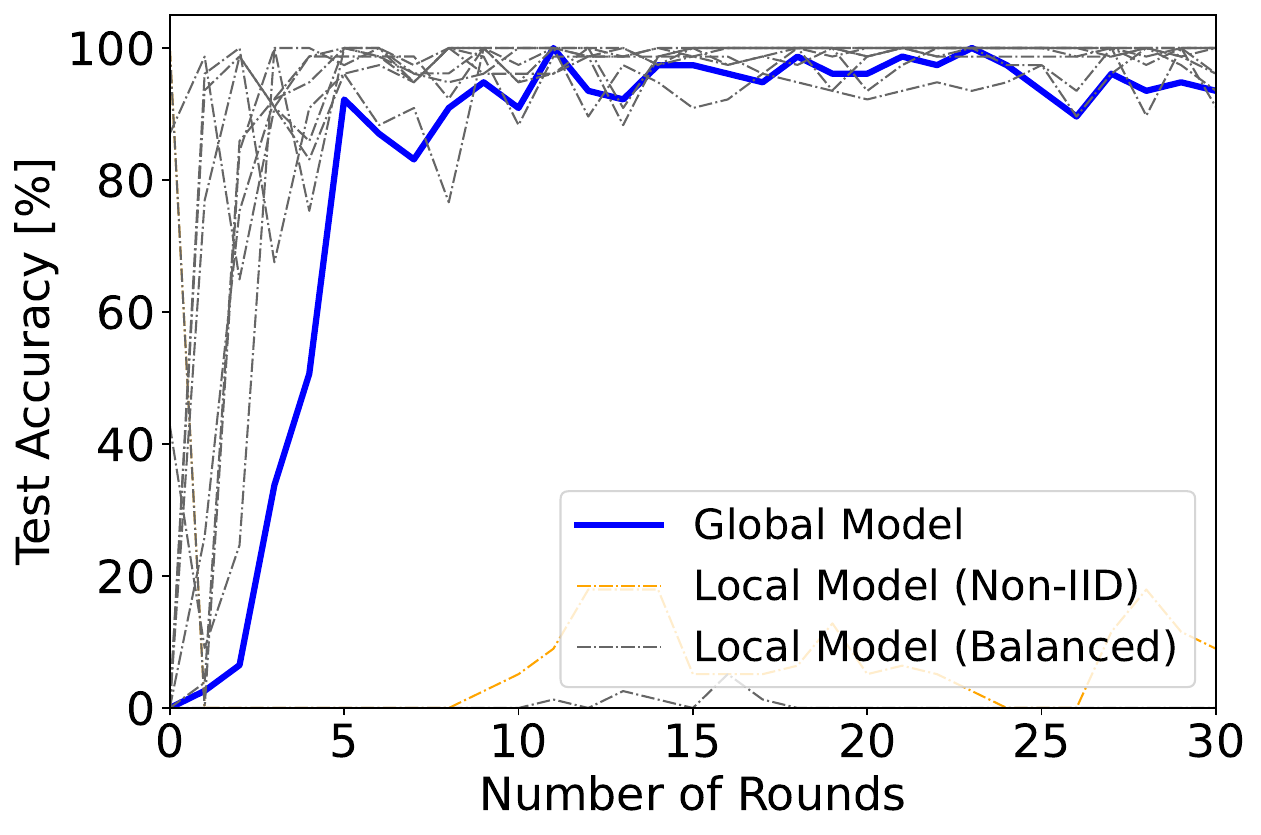} }}%
    \subfloat[\centering 70\% of  C\&C Missing Client]{{\includegraphics[height=3cm]{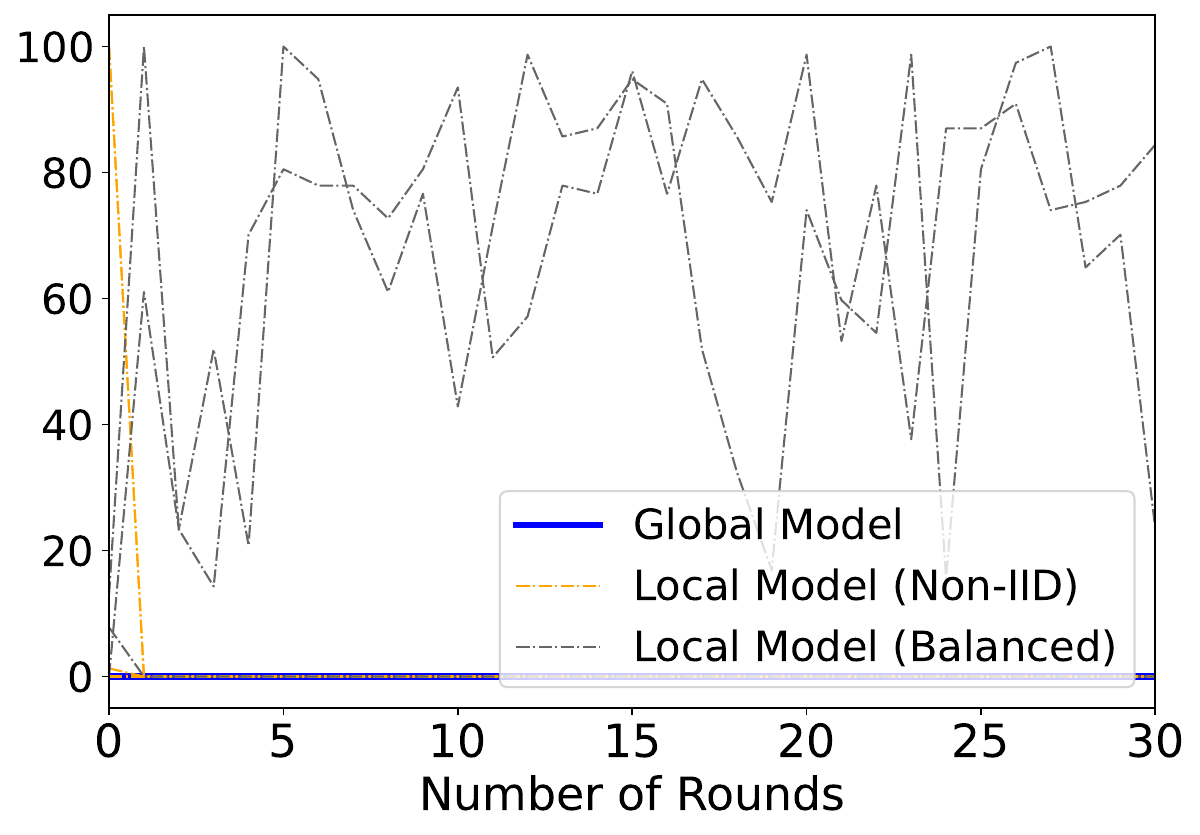} }}%    
    \caption{(a) and (b): Bdvl Test Accuracy with Absence of Rootkits Malware Family Data; (c) and (d): JAKORITAR  Test Accuracy with Absence of C\&C Malware Family Data when CyberForce with 10 Clients}
    \label{fig:bdvl_rootkits_absense}
\end{figure}

From the malware point of view, Ransomware can be effectively mitigated with FedAvg by leveraging its two MTD actions and the cooperative learning aspect. However, if none of the clients have seen Ransomware data, the likelihood of agents making the correct choice diminishes to 33.25\%. In contrast, the malware belonging to the C\&C family (The Tick, Jakoritar, and Dataleak) as well as the rootkits family (Beurk and Bdvl) exhibit an overall mitigation success rate exceeding 94\% with FedAvg, despite the number of weak non-IID clients varies. This can be attributed to the collaborative learning mechanism of the FRL architecture, where even clients lacking specific malware data can still benefit from the knowledge shared by other clients possessing that data. Another reason is that similar behaviors and MTD action decisions in the same malware family enable the agent to learn appropriate actions transferred from other malware data. Thus, further experiments are conducted to verify this hypothesis. 

\begin{figure}[t]
    \centering
    \subfloat[\centering 10\% of Rootkit Missing Client]{{\includegraphics[height=3cm]{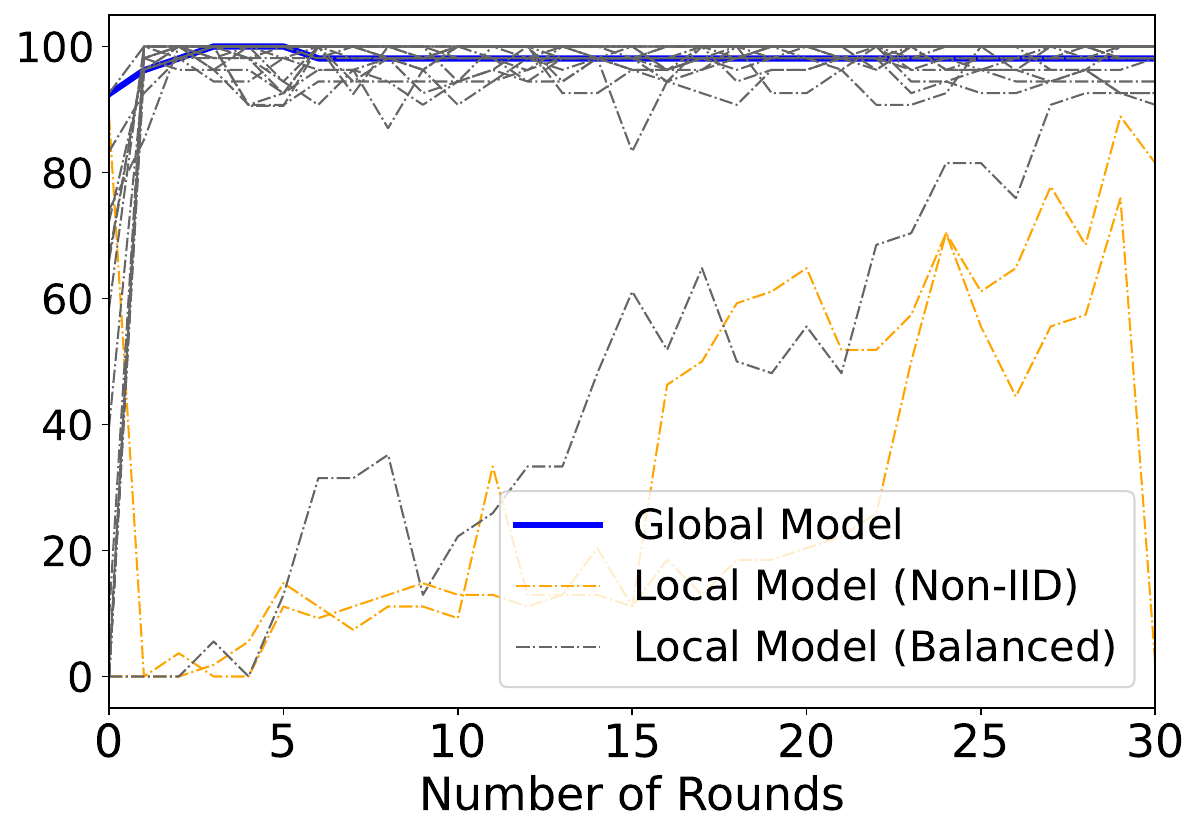} }}%
    \subfloat[\centering 70\% of  Rootkits Missing Client]{{\includegraphics[height=3cm]{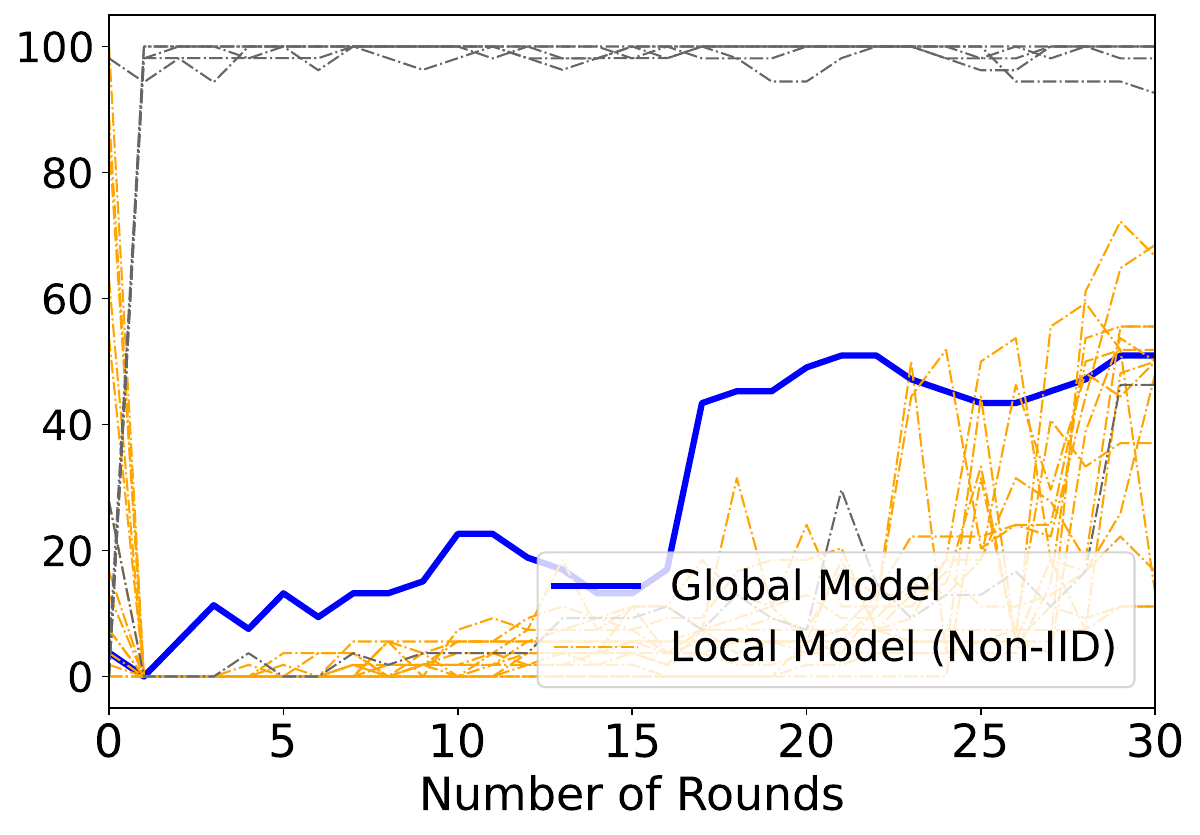} }} \\
    \subfloat[\centering 10\% of  C\&C Missing Client]{{\includegraphics[height=3cm]{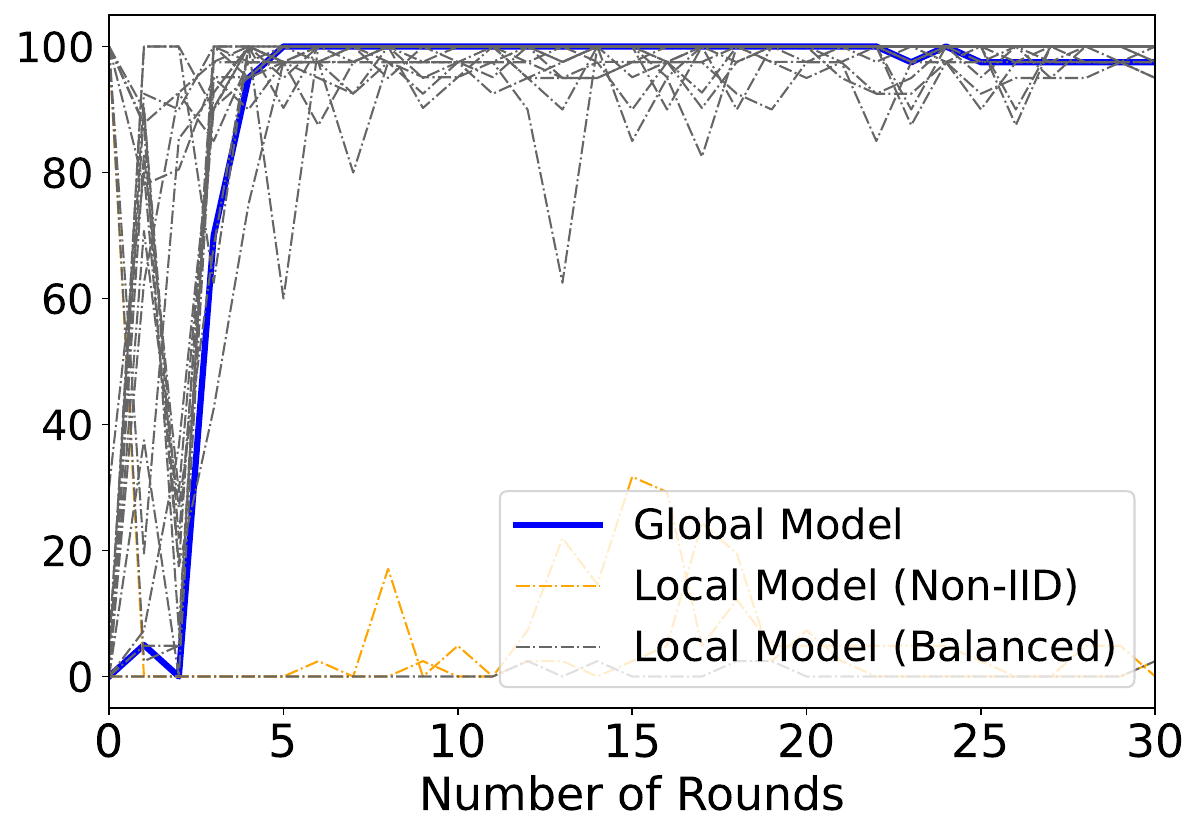} }}%
    \subfloat[\centering 70\% of  C\&C Missing Client]{{\includegraphics[height=3cm]{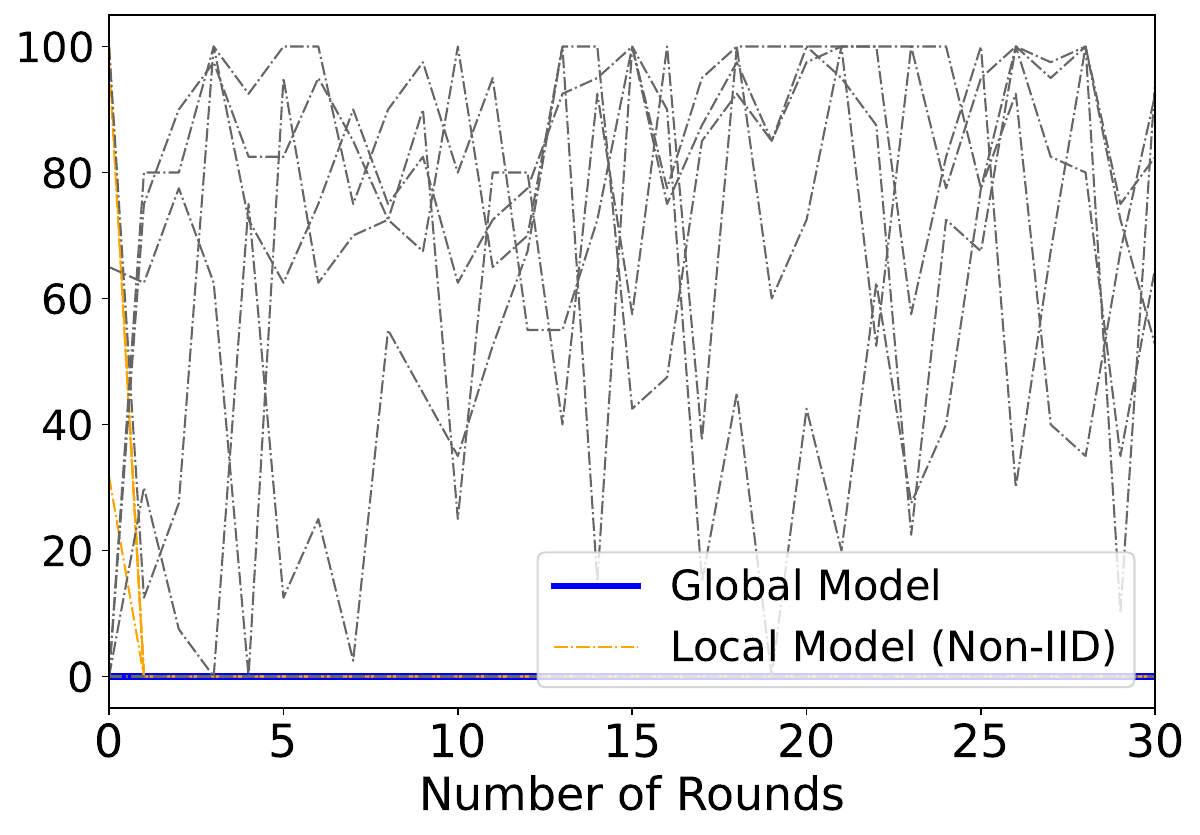} }}%    
    \caption{(a) and (b): Bdvl Test Accuracy with Absence of Rootkits Malware Family Data; (c) and (d): JAKORITAR  Test Accuracy with Absence of C\&C Malware Family Data when CyberForce with 20 Clients}
    \label{fig:20_bdvl_rootkits_absense}
\end{figure}

{Similar patterns are observed when the number of nodes within the federation is increased to 20, as illustrated in TABLE {\ref{tbl:weak_non_iid_20clients}}. When the proportion of nodes not exposed to specific malware is relatively low (below 40\%), the global model demonstrates efficacy in mitigating all forms of malware. However, as the proportion of nodes with missing data escalates to 70\%, the FedAvg and Trimmed Mean approaches prove effective only against The Tick, Dataleak, Bdvl, and Ransomware\_PoC. In this scenario, the appropriate defense strategy is no longer identifiable for Jakoritar and Beurk. Furthermore, when the percentage of nodes with missing data reaches 100\%, none of the methods, including FedAvg, Krum, or Trimmed Mean, are effective in mitigating attacks against Jakoritar, Dataleak, Bdvl, and Beurk. Nevertheless, The Tick and Ransomware\_PoC remain susceptible to effective mitigation. This suggests that knowledge transfer continues to be a viable mechanism within the FRL framework.}

To validate the transfer learning capability of CyberForce, experiments are conducted where one family of malware is not seen for one client at a time. In this experiment, FedAvg is used as the aggregation algorithm. As illustrated in \figurename~\ref{fig:bdvl_rootkits_absense}, Beurk is not seen by any client and Bdvl is missing by one client (a), and seven clients (b). In the scenario depicted in \figurename~\ref{fig:bdvl_rootkits_absense} (a), where there is only one client without rootkit data, the global model demonstrates a convergence with an accuracy exceeding 98\% for Bdvl. On the opposite, when the number of clients without rootkit data increases to seven, as shown in \figurename~\ref{fig:bdvl_rootkits_absense} (b), the global model could not converge effectively due to the absence of transfer learning from similar behaviors. Similarly, the federated global model is able to converge with an accuracy of approximately 96\% when only one client does not see the C\&C malware family data, as shown in \figurename~\ref{fig:bdvl_rootkits_absense} (c). However, as the number of clients lacking the C\&C malware data increases to seven, the accuracy of the global model significantly drops, approaching zero, as illustrated in \figurename~\ref{fig:bdvl_rootkits_absense} (d). This experiment provides an explanation for the noteworthy performance of FedAvg presented in \tablename~\ref{tbl:weak_non_iid}, as all the clients have only one absence malware data, allowing the agents to acquire the knowledge through the observation of similar behavior exhibited by the same malware family. 
{When the number of nodes in the federation is increased to 20, as illustrated in Fig.{\ref{fig:20_bdvl_rootkits_absense}}, the outcomes remain consistent with those observed at 10 nodes. In instances where a smaller proportion of nodes are missing data on some of the attacks, these agents are still able to learn how to mitigate attacks they have not seen before through the mechanism of FL's knowledge migration. However, when the proportion of nodes with data missing exceeds 70\%, the efficacy of this mechanism diminishes significantly.}

% \begin{figure}[H]
%     \centering
%     \subfloat[\centering One Non-IID Client]{{\includegraphics[height=3cm]{img/non_iid_Behavior.CNC_BACKDOOR_JAKORITAR_1_v2.pdf} }}%
%     \subfloat[\centering Seven Non-IID Client]{{\includegraphics[height=3cm]{img/non_iid_Behavior.CNC_BACKDOOR_JAKORITAR_7_v2.pdf} }}%
%     \caption{JAKORITAR Mitigation Accuracy of Non-IID Clients with Absence of C\&C Malware Family Data}
%     \label{fig:JAKORITAR_cnc_absense}
% \end{figure}

\begin{figure}[H]
    \centering
    \subfloat[\centering CyberForce with 10 Clients]{{\includegraphics[height=2.9cm]{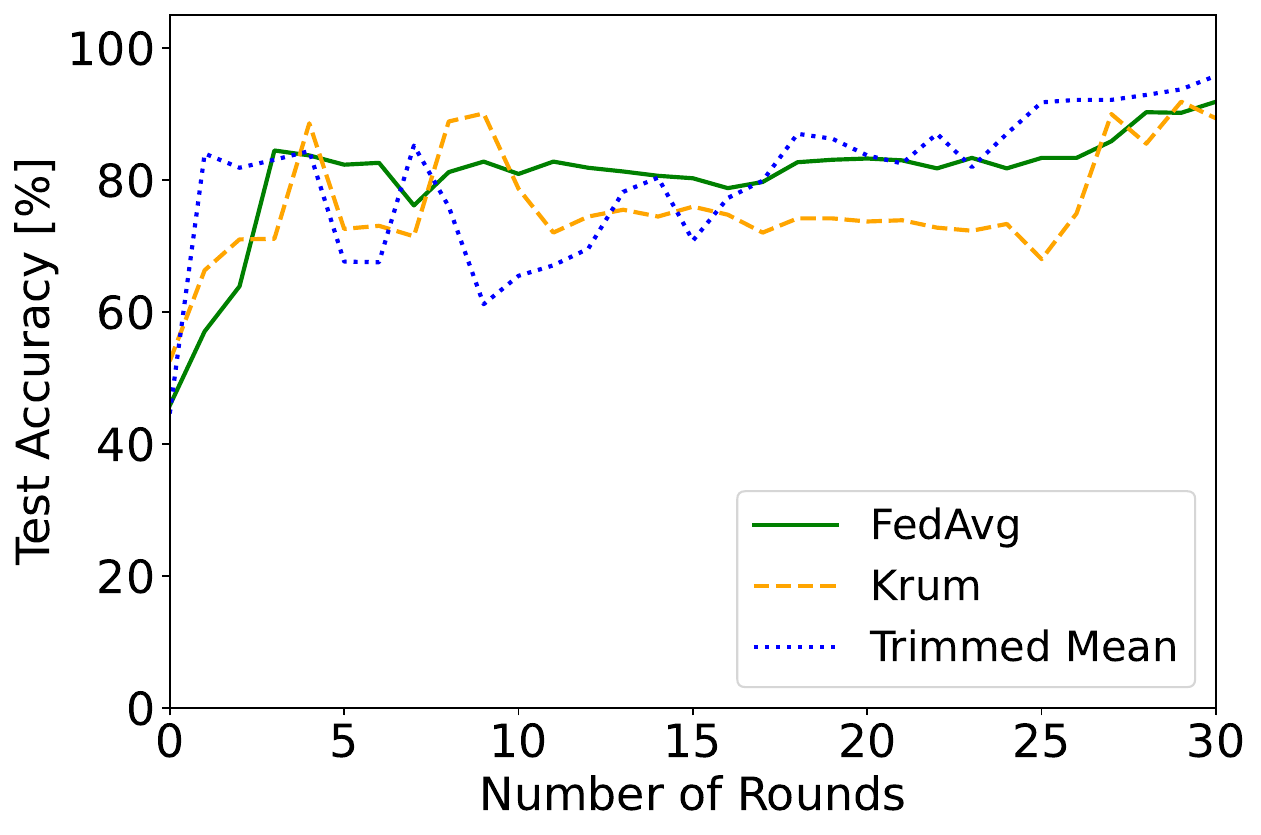} }}%
    \subfloat[\centering CyberForce with 20 Clients]{{\includegraphics[height=2.9cm]{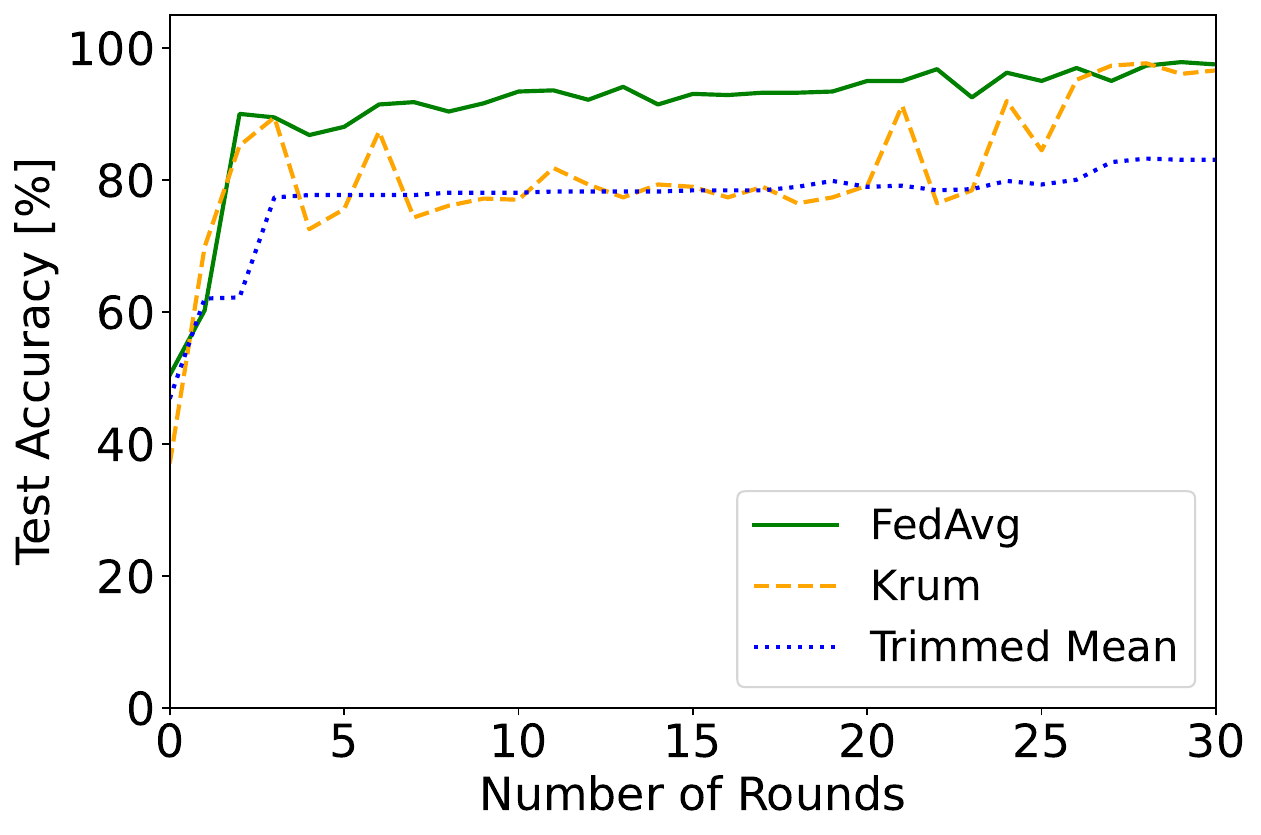} }} \\
\caption{Test Accuracy of CyberForce with Strong Non-IID}
\label{fig:strong_non_iid}
\end{figure}

In the strong non-IID scenario each client is not fully exposed to all six malware attacks. In this setup, three varieties of malware are randomly absent on each client. Compared to the weak non-IID scenarios (\tablename~\ref{tbl:weak_non_iid}), the results of strong non-IID demonstrate a significant decline in accuracy in all three aggregation algorithms, fluctuating between 60\% and 85\% {for both of the federations with 10 and 20 clients}, as illustrated in \figurename~\ref{fig:strong_non_iid}, which demonstrates the constraints of transfer learning and the collaborative learning capabilities of the FRL framework. {Meanwhile, FedAvg exhibited enhanced stability, indicating its superior capacity to address challenges associated with non-IID data distributions.}

% \begin{figure}[h]
% \centering
% \includegraphics[width=0.8\linewidth]{img/strong-non-iid.pdf}
% \caption{Test Accuracy of CyberForce with Strong Non-IID}
% \label{fig:strong_non_iid}
% \end{figure}

% To conclude, it is clear that a CyberForce client is able to benefit from other clients as well as the same family of malware data through collaborative learning and transfer learning in spite of the non-IID environment. Moreover, FedAvg has a better performance against non-IID compared to Krum and Trimmed Mean. 

\subsection{\textbf{Experiment 3}: Robustness Analysis of CyberForce}

{The experiments previously discussed illustrate that the proposed FRL framework is capable of effectively mitigating various types of malware in a distributed environment. The aim of Experiment 3 is to assess the robustness of the FRL framework in the face of poisoning attacks. In the design of this experiment, a threat model focused on internal attacks within FL is employed {\cite{kumar2023impact}}, which is analyzed in this paper concerning the objective and capabilities of the internal attacker.

\textbf{Internal Attacker's Objective:} The distributed nature of FL presents challenges regarding the integrity of the local models submitted by local nodes. In an insecure FL environment, there is no assurance that all participating nodes will contribute honest local models. Malicious nodes may submit compromised models to the central server by engaging in data poisoning attacks, which involve altering the input data, or model poisoning attacks, which entail manipulating the model itself. Such actions can undermine the accuracy of the global model. Thus, the objective of the attacker is to disrupt the convergence of the global model, ultimately impairing its capacity to select the appropriate MTD strategy.

\textbf{Internal Attacker's Capabilities:} Given that the attacker is internal, functioning as clients within the distributed system, the poisoning attacks designed in this paper are characterized as black-box attacks. In this context, the attacker has the capability to manipulate and interfere with local data and models; however, direct alterations to the global model are not feasible. Furthermore, the attacker may exert control over multiple compromised nodes, thereby influencing the poisoned nodes ratio (PNR), and can execute cooperative attacks. Notably, the attacker applies the same operation across all compromised nodes.
}

\subsubsection{{Experimental Settings}}
{A commonly used methodology for evaluating the robustness of FL models is to assess their resistance in the face of poisoning attacks {\cite{kumar2023impact}} {\cite{feng2024dart}}. TABLE {\ref{tab:exp3_config}} shows the setup of Experiment 3 using three different types of poisoning attacks, including Label Flipping, Sample Poisoning, and Model Poisoning.}

In particular, the following {three} attacks have been designed and implemented: (i) {Label Flipping} attack, in which all normal data labels are manipulated as malware labels and all malware labels are randomly changed to normal labels, with the aim of undermining the effectiveness of the AD system in delivering undesirable rewards; (ii) {Sample Poisoning attack, which involves the deliberate introduction of 50\% Gaussian noise into the training data of malicious nodes, thereby corrupting the training process of the models. The objective of this attack is to diminish the overall performance of the global model by disseminating these compromised models to the aggregator;} (iii) Model poisoning attack, where 50\% Gaussian noise is injected into the neural network after each round of training, thus affecting the performance of the agent. {In terms of attacked nodes, this experiment considers an increase in the poisoned nodes ratio from 10\% to 90\% to test the limits of each robustness aggregation function. The other experimental settings are the same as in Experiment 2.}

\begin{table}[h]
\centering
\caption{Configuration of Experiment III}
\label{tab:exp3_config}
\resizebox{\columnwidth}{!}{%
\begin{tabular}{lll}
\toprule
\multicolumn{2}{l}{\textbf{CONFIGURATION}} & \textbf{VALUE} \\ \midrule
\multirow{5}{*}{\textbf{Federation}} & Number of Clients & 10, 20 \\ \cmidrule(l){2-3}
 & Total Rounds & 30 \\ \cmidrule(l){2-3}
 & Episodes in Each Round & 100 \\ \cmidrule(l){2-3}
 & Aggregation Function& \textit{FedAvg}, \textit{Krum}, \textit{Trimmed Mean} \\ \hline 
\multirow{3}{*}{\textbf{Robustness}} & Poisoning Attack & \begin{tabular}[c]{@{}l@{}}Label Flipping\\ Sample Poisoning \\ Model Poisoning \end{tabular} \\ \cmidrule(l){2-3}
  & Poisoned Nodes Ratio (PNR) & 10\%, 30\%, 50\%, 70\%, 90\% \\ \hline 
\end{tabular} 
}
\end{table}

\subsubsection{Robustness Analysis}
\begin{figure}[t]
    \centering
    \subfloat[\centering Label Flipping (10 Clients)]{{\includegraphics[height=3cm]{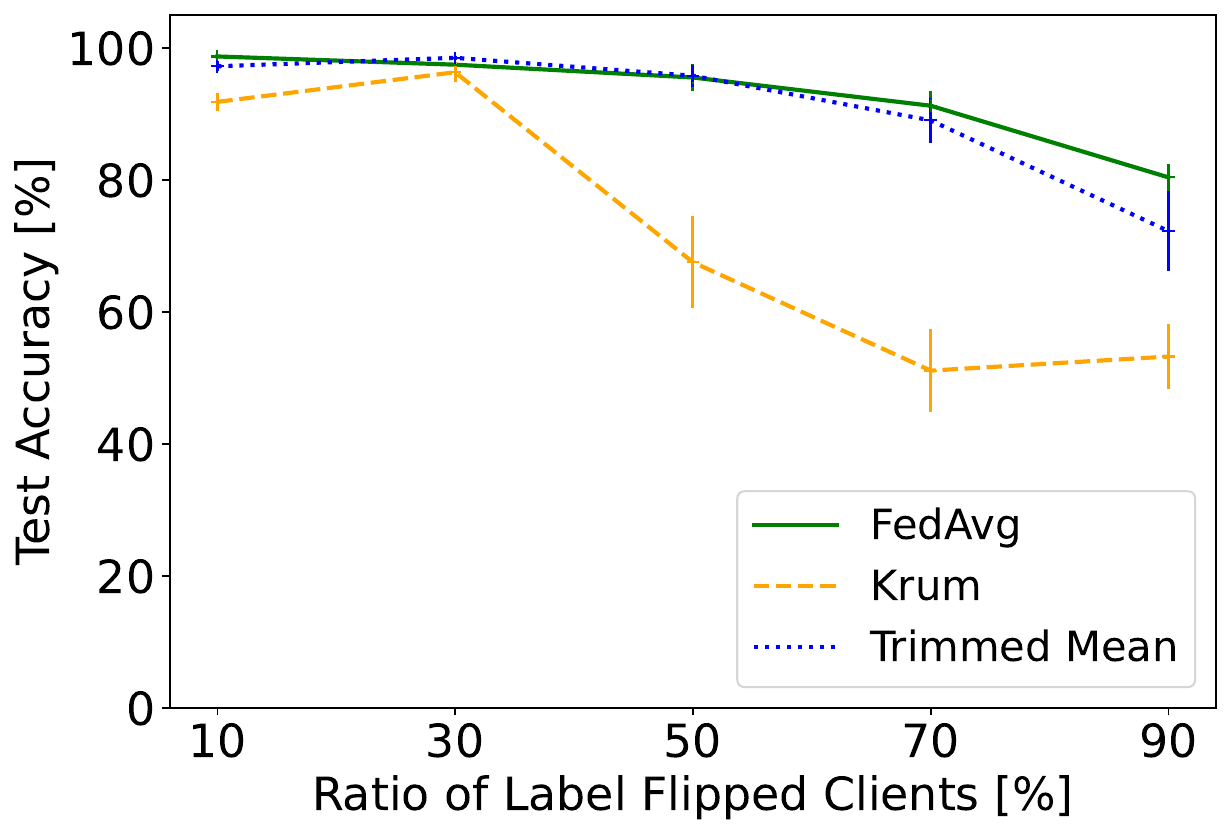} }}%
    \subfloat[\centering Label Flipping (20 Clients)]{{\includegraphics[height=3cm]{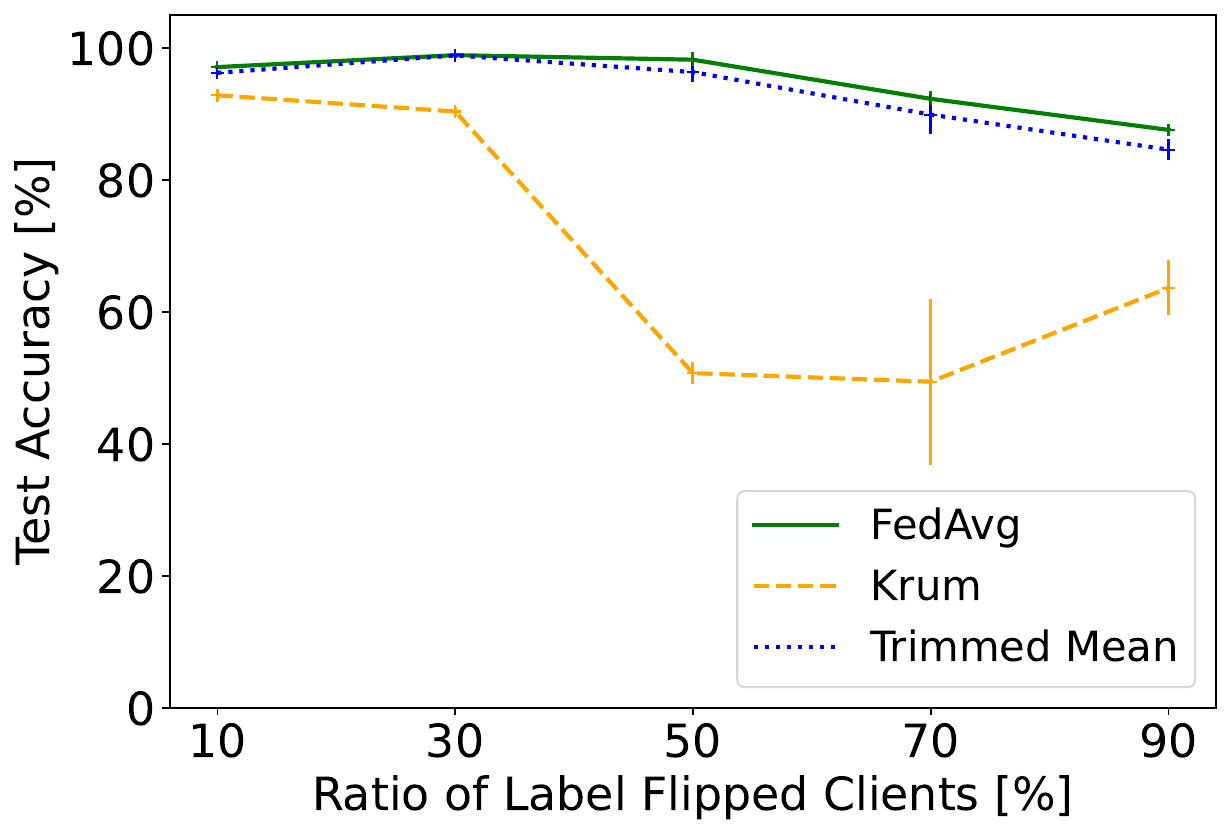} }} \\
    \subfloat[\centering Sample Poisoning (10 Clients)]{{\includegraphics[height=3cm]{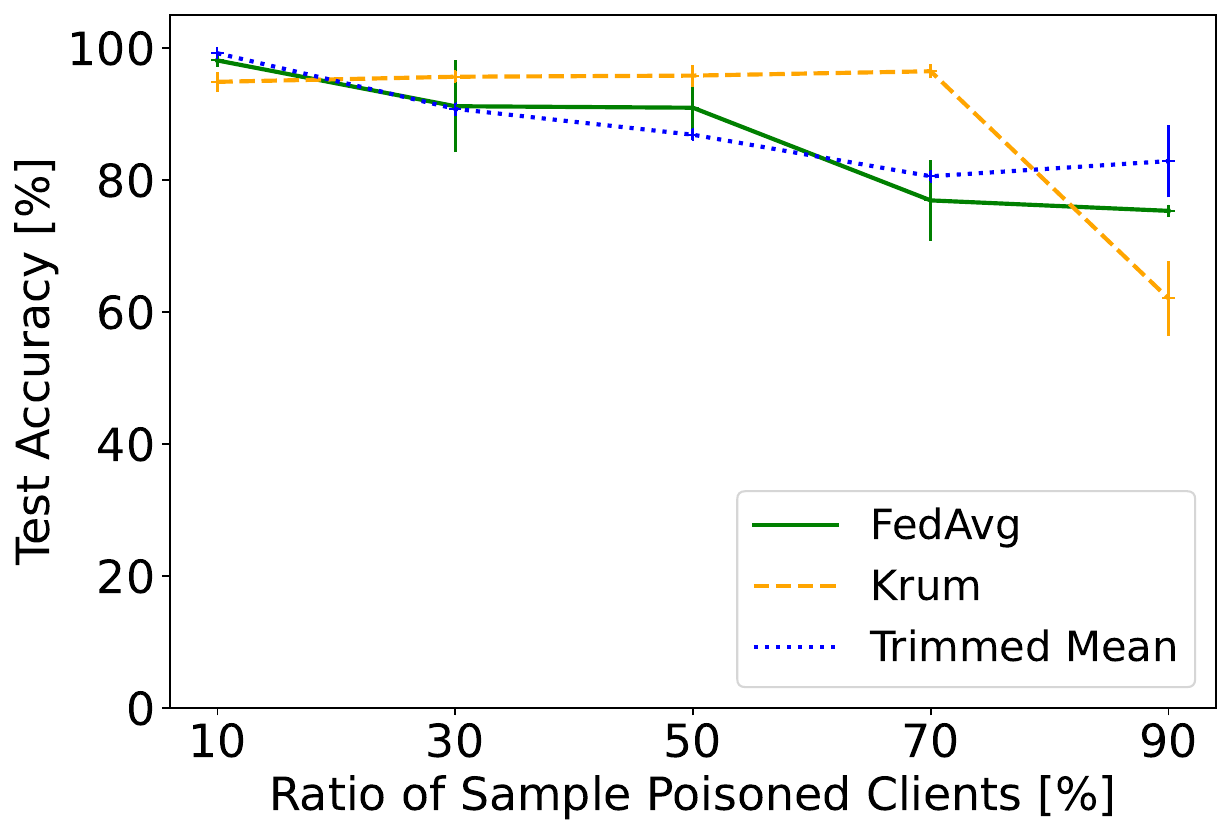} }}%
    \subfloat[\centering Sample Poisoning (20 Clients)]{{\includegraphics[height=3cm]{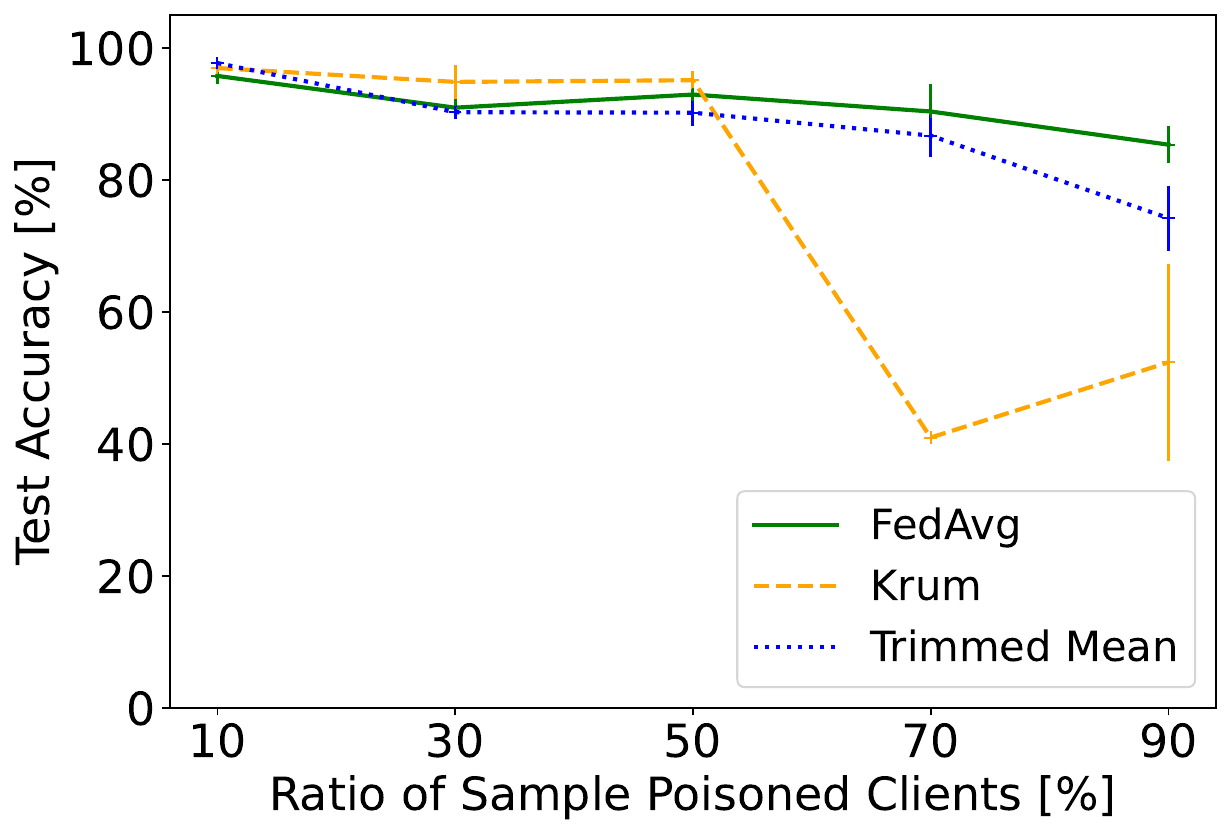} }} \\
    \subfloat[\centering Model Poisoning (10 Clients)]{{\includegraphics[height=3cm]{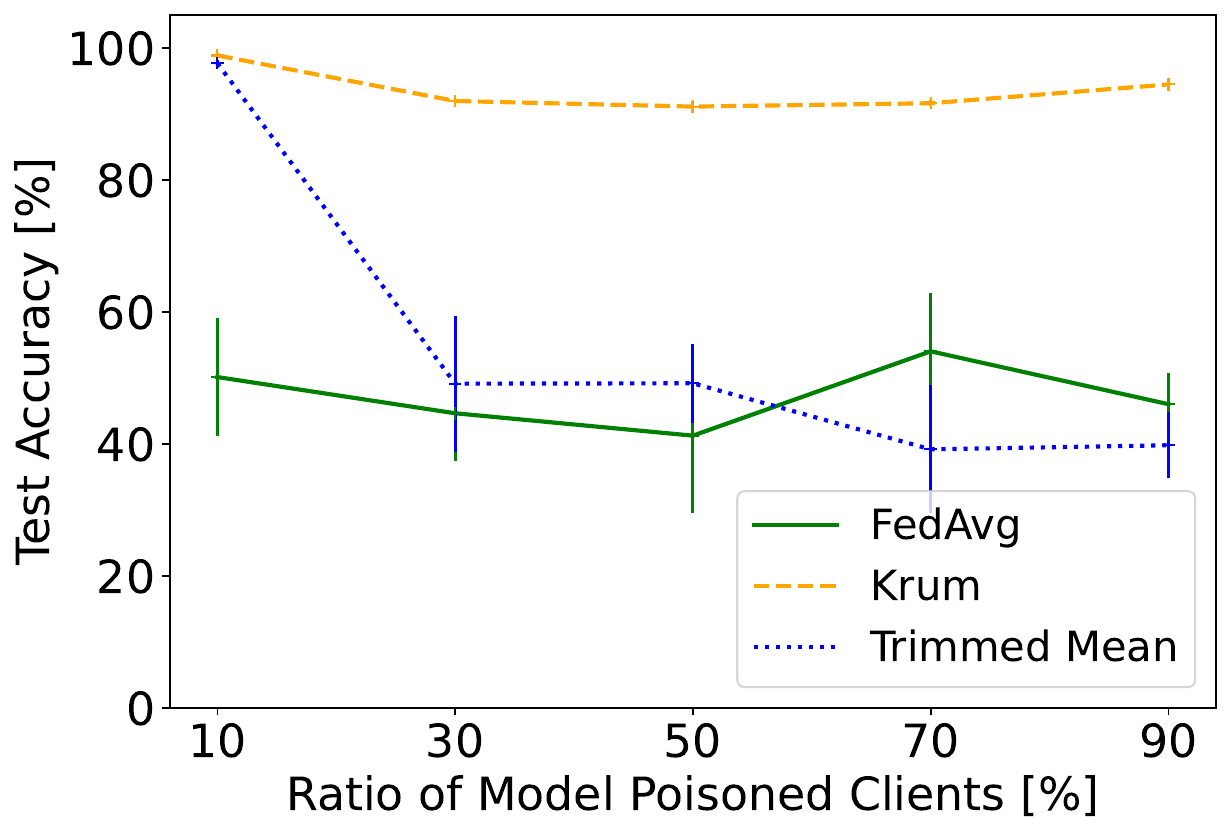} }}%
    \subfloat[\centering Model Poisoning (20 Clients)]{{\includegraphics[height=3cm]{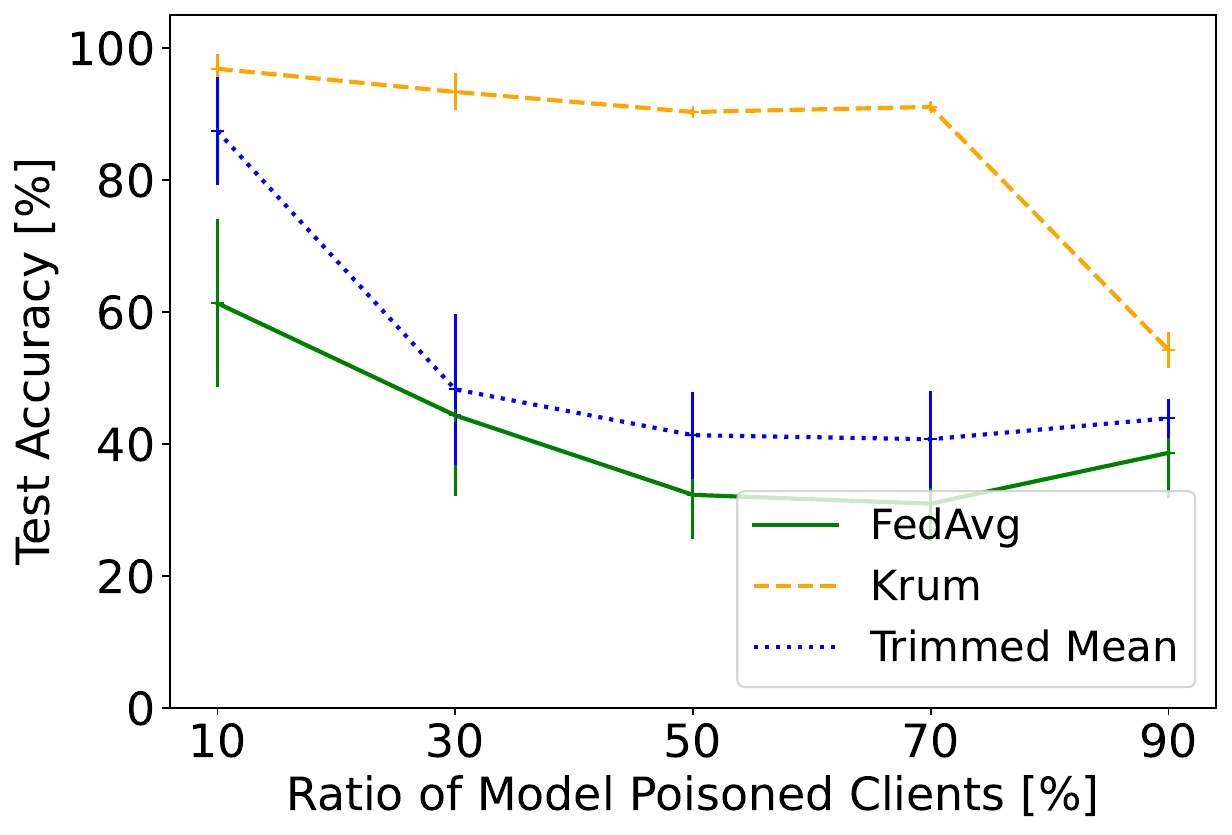} }}%    
    \caption{Test Accuracy of Different Aggregation Algorithms with Malicious Attacks}
    \label{fig:data_and_model_poisoning}
\end{figure}

By progressively increasing the number of {label flipped} clients, the accuracy of all three aggregation algorithms declines, as illustrated in \figurename~\ref{fig:data_and_model_poisoning} (a). FedAvg and Trimmed Mean showed better resistance to {label flipping} attacks, with their accuracy starting to decline only when the share of poisoned clients is greater than half. However, Krum performs the worst in model poisoning, and its accuracy drops significantly to 60\% when the number of poisoned clients is greater than three. {When the number of nodes in the federation is increased to 20, as shown in Fig.{\ref{fig:data_and_model_poisoning}} (b), the same results as with 10 nodes are obtained. This suggests that if AD fails to provide the correct state, the agent cannot learn how to choose the correct strategy to mitigate different malware. Thus, the accuracy of AD in yielding correct outcomes is essential for a reactive MTD system.}

{The results of the experiments are shown in Figs. {\ref{fig:data_and_model_poisoning}} (c) and (d), which illustrate the performance of the three aggregation functions at both 10 and 20 nodes. The results indicate that these functions exhibit considerable resilience against attacks when the proportion of malicious nodes is below 50\%. However, as the proportion of malicious nodes exceeds 50\%, there is a notable decline in the system's effectiveness across all three aggregation functions. This decline is particularly noticeable in Krum, which is less than 50\% accurate when the percentage of malicious nodes reaches 70\%. From an adversarial perspective, it is evident that manipulating the training data can significantly impair the agents' capacity to select the correct strategies. Conversely, with respect to defensive strategies, both FedAvg and Trimmed Mean demonstrate superior efficacy in mitigating the impact of sample poisoning attacks.}
% \begin{figure}[h]
%     \centering
%     \subfloat[\centering Data Poisoning Attacks]{{\includegraphics[height=4.5cm]{img/data_poisoning_accuracy_3aggs.pdf} }}\\%
%     \subfloat[\centering Model Poisoning Attacks]{{\includegraphics[height=4.5cm]{img/model_poisoning_accuracy_3aggs.pdf} }}%    
%     \caption{ Test Accuracy of Different Aggregation Algorithms with Malicious Attacks}
%     \label{fig:data_and_model_poisoning}
% \end{figure}

In contrast, Krum performs the best when confronting model poisoning attacks, as shown in \figurename~\ref{fig:data_and_model_poisoning} {(e) and (f)}, maintaining an accuracy of over 95\% even when 70\% of the clients have been poisoned. It is observed that the Euclidean distance between poisoned clients and the rest of the clients is consistently greater than the distance between benign clients. Consequently, the Krum algorithm can detect and choose the benign model as the global model. FedAvg exhibits the least resilience against model poisoning attacks, as evidenced by a significant decrease in its overall accuracy to nearly 50\% when only one client has been poisoned. Trimmed Mean is effective in mitigating low-level mod poisoning attacks, specifically, those involving only one poisoned client. Nevertheless, when the number of affected clients surpasses three, the performance of Trimmed Mean aligns with that of the FedAvg method.

\begin{figure*}[t]
    \centering
    \subfloat[\centering Label Flipping and Benign]{{\includegraphics[height=4cm]{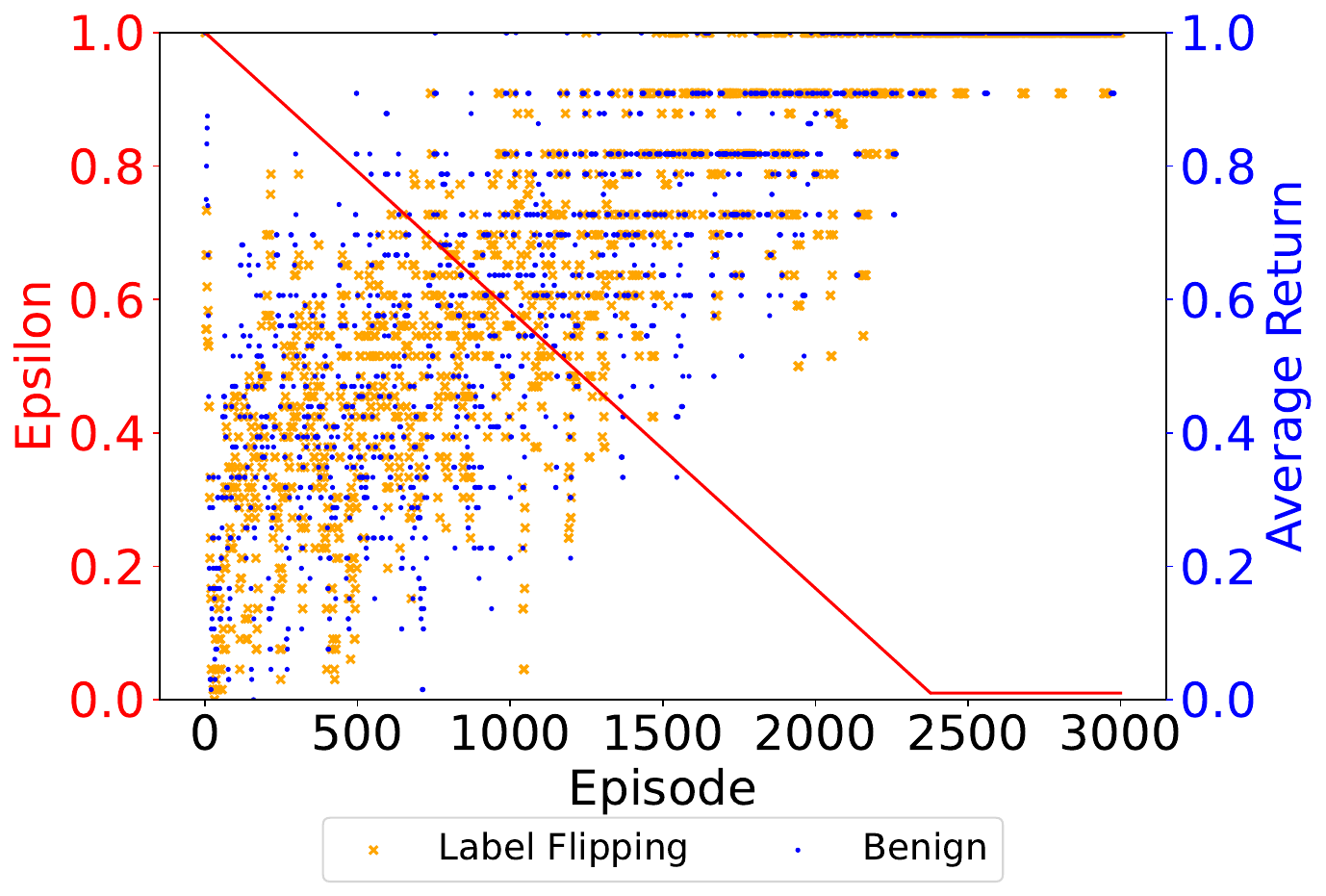} }}%
    \subfloat[\centering Sample Poisoning and Benign]{{\includegraphics[height=4cm]{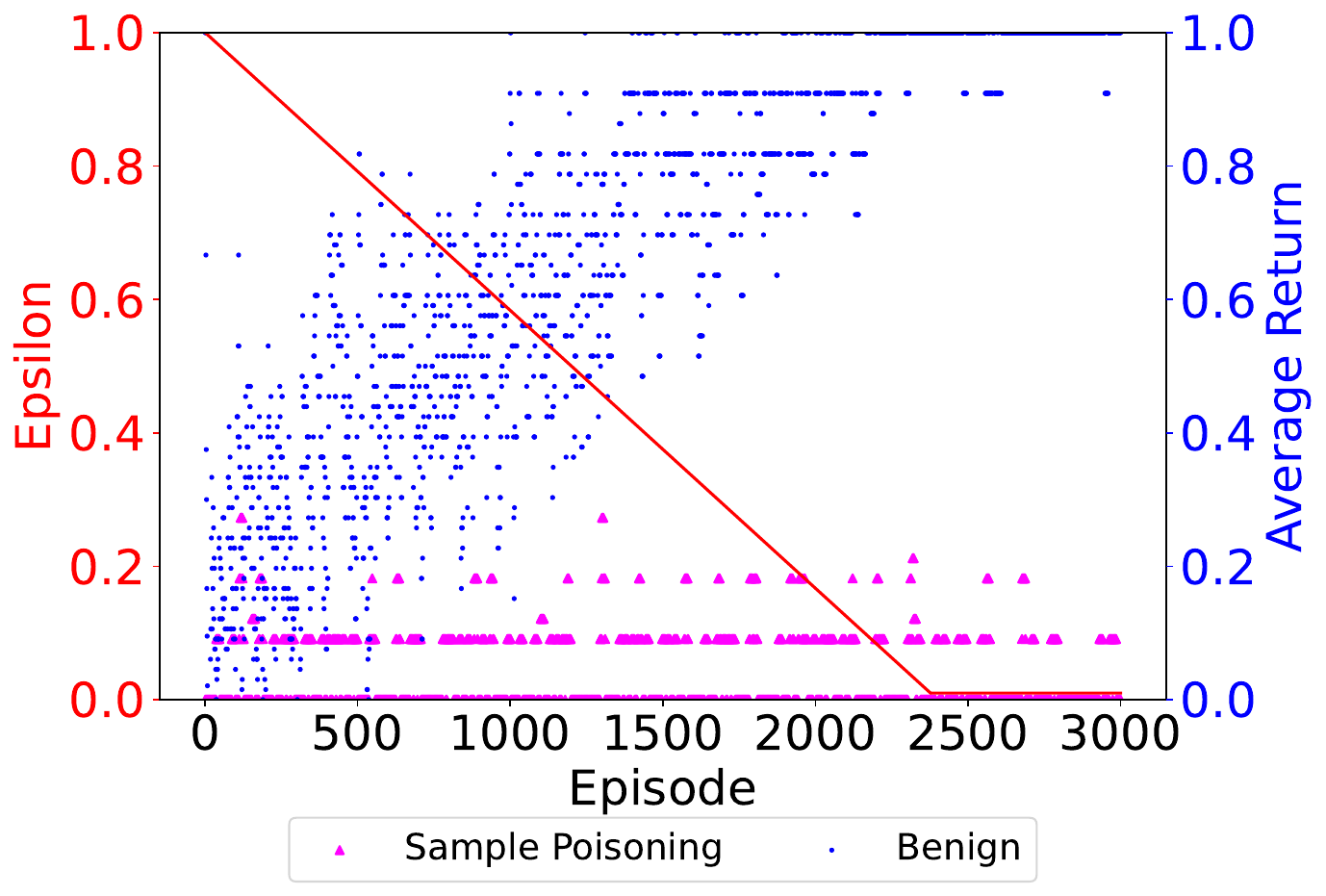} }} 
    \subfloat[\centering Model Poisoning and Benign]{{\includegraphics[height=4cm]{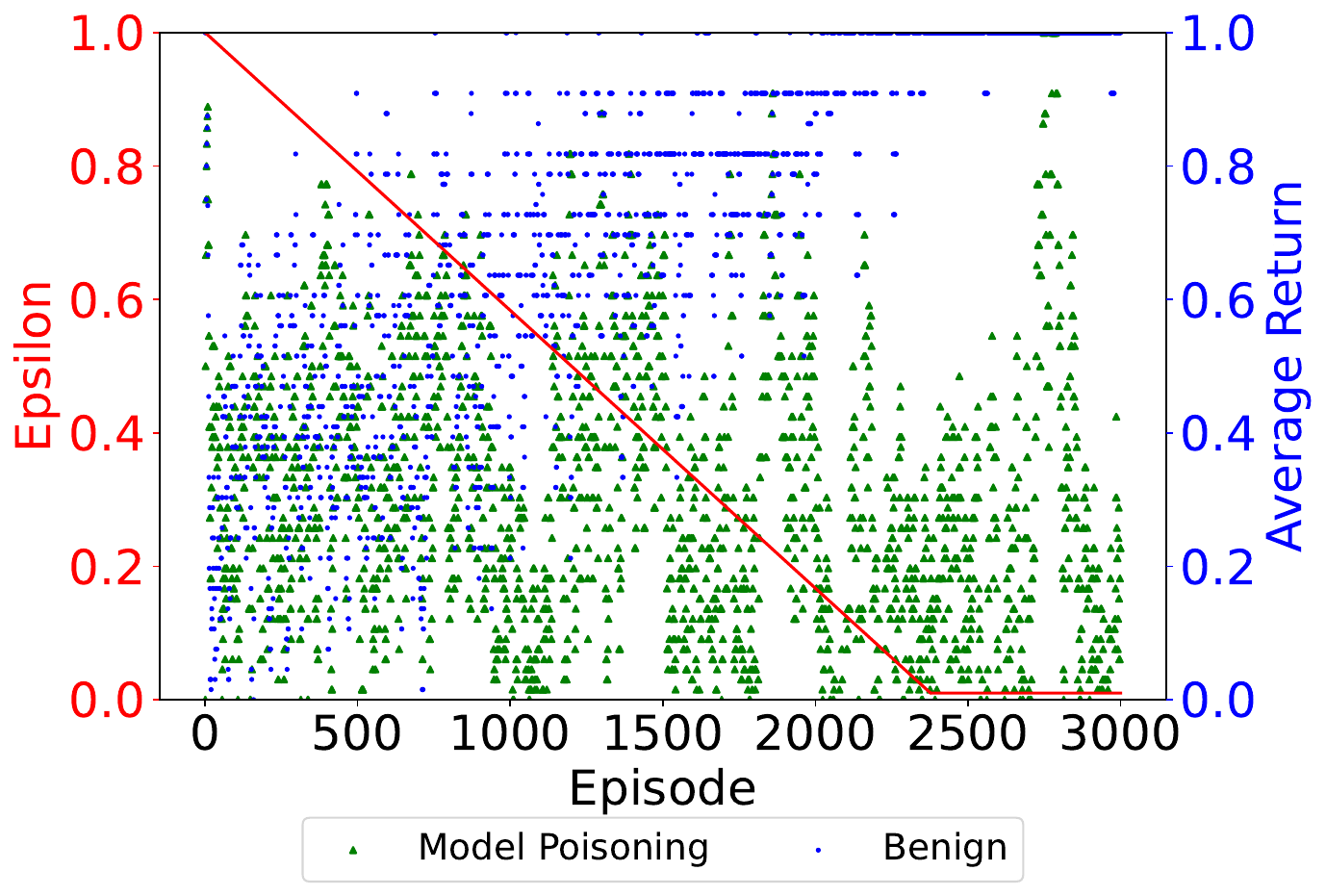} }}%   
    \caption{Learning Curve in FedAvg with Malicious Attacks}
    \label{fig:malicious_attacks}
\end{figure*}

% \begin{figure}[h]
%     \centering
%     \subfloat[\centering Model Poisoned and Benign]{{\includegraphics[height=5cm]{img/client_learning_curve_mp.pdf} }}\\%
%     \subfloat[\centering Data Poisoned and Benign]{{\includegraphics[height=5cm]{img/client_learning_curve_dp.pdf} }}%    
%         \caption{Learning Curve in FedAvg with Malicious Attacks}
%     \label{fig:malicious_attacks}
% \end{figure}

While closely inspecting FedAvg, it can be noticed that when exposed to model poisoning attacks, the learning curve of the poisoned client is completely destroyed, as shown in \figurename~\ref{fig:malicious_attacks} (c), which indicates that the client is unable to learn any meaningful information. Nevertheless, FedAvg presents resilience against {label flipping} attacks, as shown in \figurename~\ref{fig:malicious_attacks} (a), where malicious agents exhibit learning curves comparable to benign agents. {In the context of sample poisoning attacks, as shown in Fig. {\ref{fig:malicious_attacks}} (b), the behavior data of the malware is similar to the normal behavior data. As a result, the agent cannot get effective rewards and learn how to make effective decisions.}

\begin{figure}[h]
    \centering
    \subfloat[\centering One Malicious Client]{{\includegraphics[height=3.8cm,trim={0 0 4cm 0},clip]{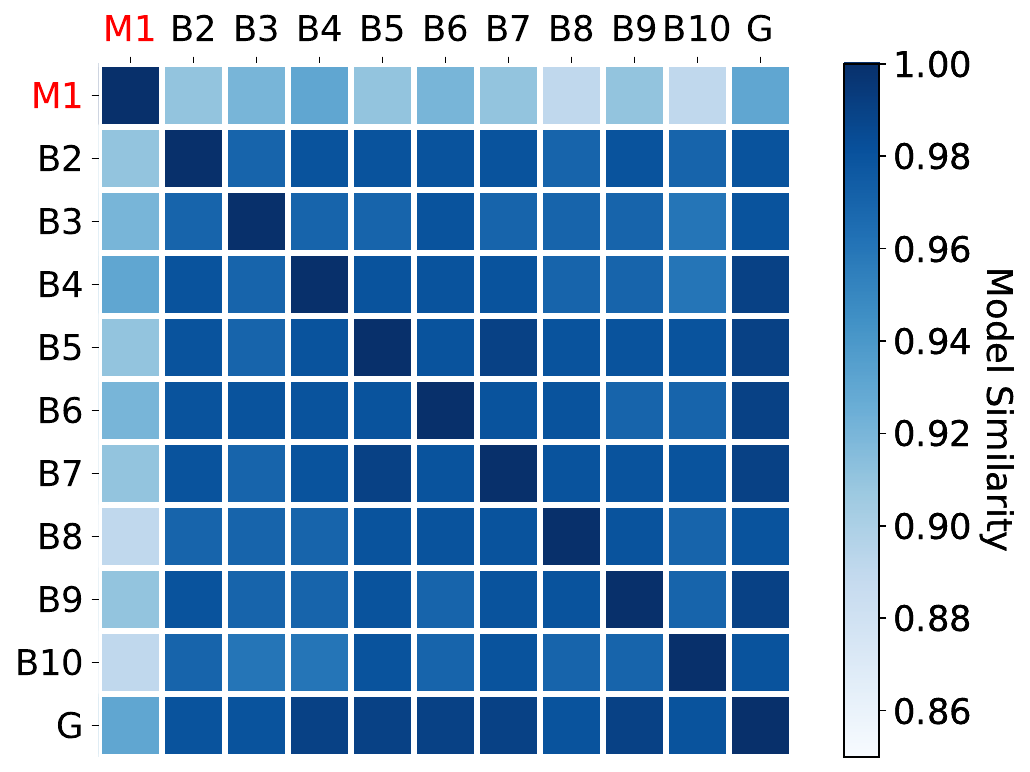} }}%
    \subfloat[\centering Five Malicious Clients]{{\includegraphics[height=3.8cm]{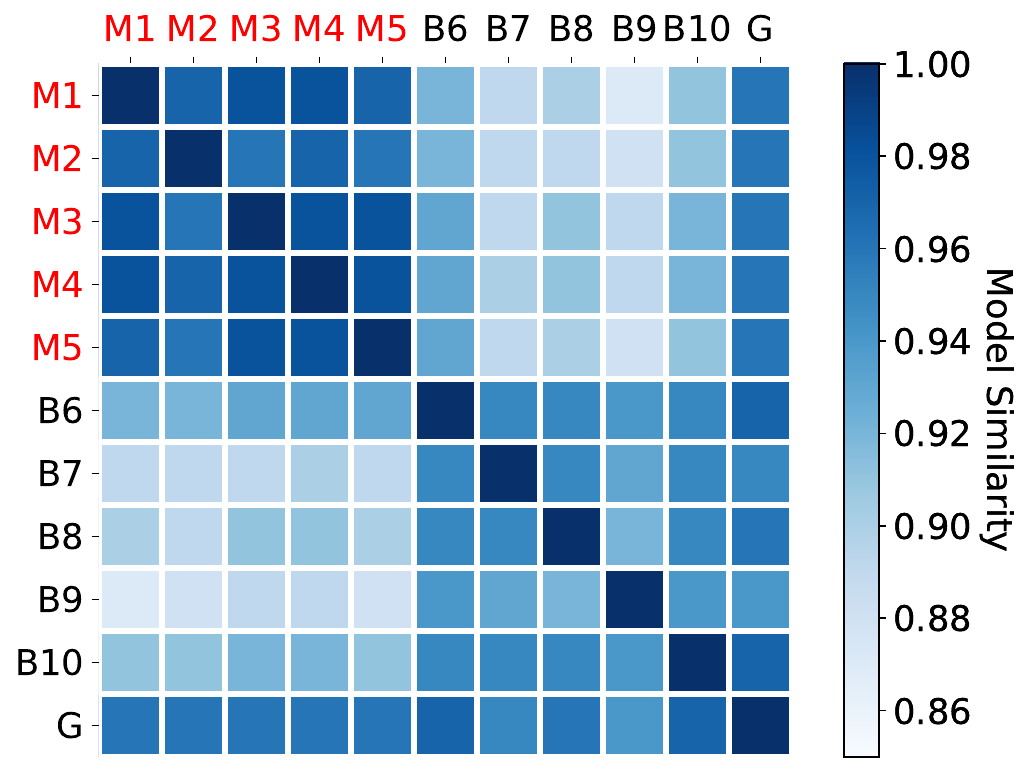}}} \\%   
    {\footnotesize Malicious Model (\textcolor{red}{M\textit{i}}), Benign Model (B\textit{j}), Global Model (G)}         \caption{Model Similarity in FedAvg with Label Flipping}
    \label{fig:datapoison_modelsim}    
\end{figure}

\figurename~\ref{fig:datapoison_modelsim} portrays the model similarity between different agents under the Label Flipping attack. When there is only one being attacked, as shown in \figurename~\ref{fig:datapoison_modelsim} (a), the global model is closer to the benign models. Whereas, when malicious clients increase to 5, the global model gets closer to the malicious model, as shown in \figurename~\ref{fig:datapoison_modelsim} (b), the accuracy of the global model starts to decrease. {This explains why with FedAvg and Trimmed Mean aggregation functions, it takes more than 50\% of the attacked nodes to make the global model's performance start to drop. Correspondingly, for Krum, when the proportion of attacked nodes exceeds 50\%, there is a high probability that an attacked model will be chosen as the global model, and therefore its effectiveness drops significantly.}

In conclusion, in IID scenarios, CyberForce significantly optimizes training time while maintaining over 98\% accuracy, outperforming centralized RL-based solutions. Its adaptability shines with aggregation algorithms like Krum, FedAvg, and Trimmed Mean, addressing diverse cyberattack challenges. Krum strengthens resilience against model poisoning attacks in IID environments, while FedAvg enhances non-IID data performance. Trimmed Mean offers balanced performance, effectively managing limited non-IID conditions and resisting minor malicious attacks.

\section{Conclusions and Future Work}
\label{sec:conclusion}

This work presents CyberForce, an FRL-based framework able to select and deploy MTD mechanisms mitigating diverse zero-day attacks on IoT devices. CyberForce incorporates behavioral fingerprinting and ML-based AD methods to identify zero-day attacks. Furthermore, it employs a federated agent that utilizes Deep-Q Learning to learn the most effective MTD technique per attack. The framework effectiveness was evaluated by deploying it on ten Raspberry Pi 4 devices, which acted as sensors of an IoT crowdsensing platform named Electrosense. Each device was attacked by six distinct malware samples (from the ransomware, C\&C, and rootkit families), and four heterogeneous and existing MTD mechanisms were considered to mitigate them. A series of experiments were conducted to assess the CyberForce selection performance, learning time, and robustness against attacks. With the aim of showcasing the suitability of CyberForce in diverse scenarios, the experiments encompassed the previous malware affecting multiple devices and involved varying data distributions (ranging from IID to non-IID). The results showed that in the IID scenario, CyberForce significantly reduces training time/episodes of existing centralized RL-based solutions by two-thirds while maintaining an accuracy rate of over 98\%. The CyberForce framework offers remarkable adaptability and flexibility by providing various aggregation algorithms such as Krum, FedAvg, and Trimmed Mean to address diverse cyberattack challenges. In scenarios involving substantial model poisoning attacks but an IID environment, the inclusion of Krum can bolster the resilience of the system. Conversely, when confronted with a non-IID situation within a secure setting, FedAvg can enhance performance by leveraging the collaborative mechanism. Meanwhile, Trimmed Mean provides a balanced performance. It is capable of effectively managing scenarios with limited non-IID conditions, while also displaying resilience against minor malicious attacks.

% In the non-IID scenario, CyberForce achieved satisfactory accuracy due to the collaborative and transferable learning nature of FL. Finally, the framework also demonstrated being robust against data poisoning attacks but was incapable of defending against model poisoning attacks.

As future work, it is planned to extend the deployment of the CyberForce framework to additional device types and diverse scenarios in order to evaluate its efficacy and scalability. Furthermore, it is intended to augment the framework and its evaluation by incorporating additional mitigation mechanisms, as well as new malware families such as cryptominers, infostealers, and botnets. Besides, the exploration of a combined approach involving vertical FL and RL is being considered.

\bibliographystyle{IEEEtran}
\bibliography{bare_jrnl}

 % \vskip -3\baselineskip
 \begin{IEEEbiography}[{\includegraphics[width=1in,clip]{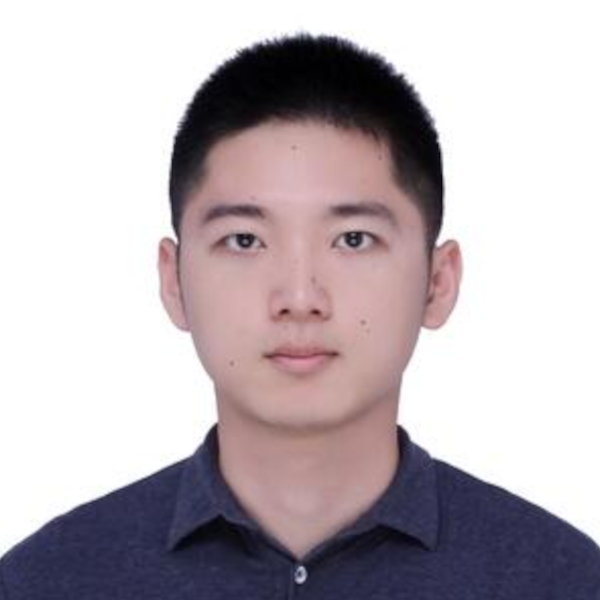}}]{Chao Feng} received the MSc degree in Informatics from the University of Zurich, Switzerland. He is currently pursuing his Ph.D. in computer science at the Communication Systems Group, Department of Informatics at the University of Zurich. His scientific interests include IoT, cybersecurity, data privacy, machine learning, and computer networks.
 \end{IEEEbiography}

 % \vskip -1\baselineskip

 \begin{IEEEbiography}[{\includegraphics[width=1in,clip]{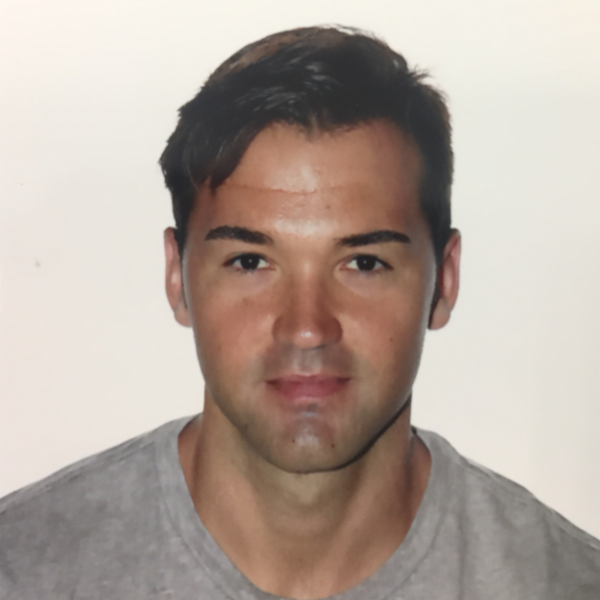}}]{Alberto Huertas Celdrán} is senior researcher at the Communication Systems Group CSG, Department of Informatics IfI, University of Zurich UZH. He received the MSc and PhD degrees in Computer Science from the University of Murcia, Spain. His scientific interests include cybersecurity, machine and deep learning, continuous authentication, and computer networks.
 \end{IEEEbiography}

 % \vskip -1\baselineskip

 \begin{IEEEbiography}[{\includegraphics[width=1in,clip]{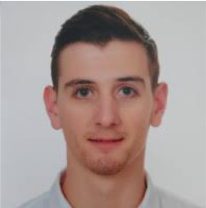}}]{Pedro M. Sánchez Sánchez} is pursuing his PhD in computer science at the University of Murcia. He received the MSc degree in Computer Science from the University of Murcia, Spain. His research interests focus on continuous authentication, networks, 5G, cybersecurity, and machine learning and deep learning.
 \end{IEEEbiography}

 % \vskip -1\baselineskip

 \begin{IEEEbiography}[{\includegraphics[width=1in,clip]{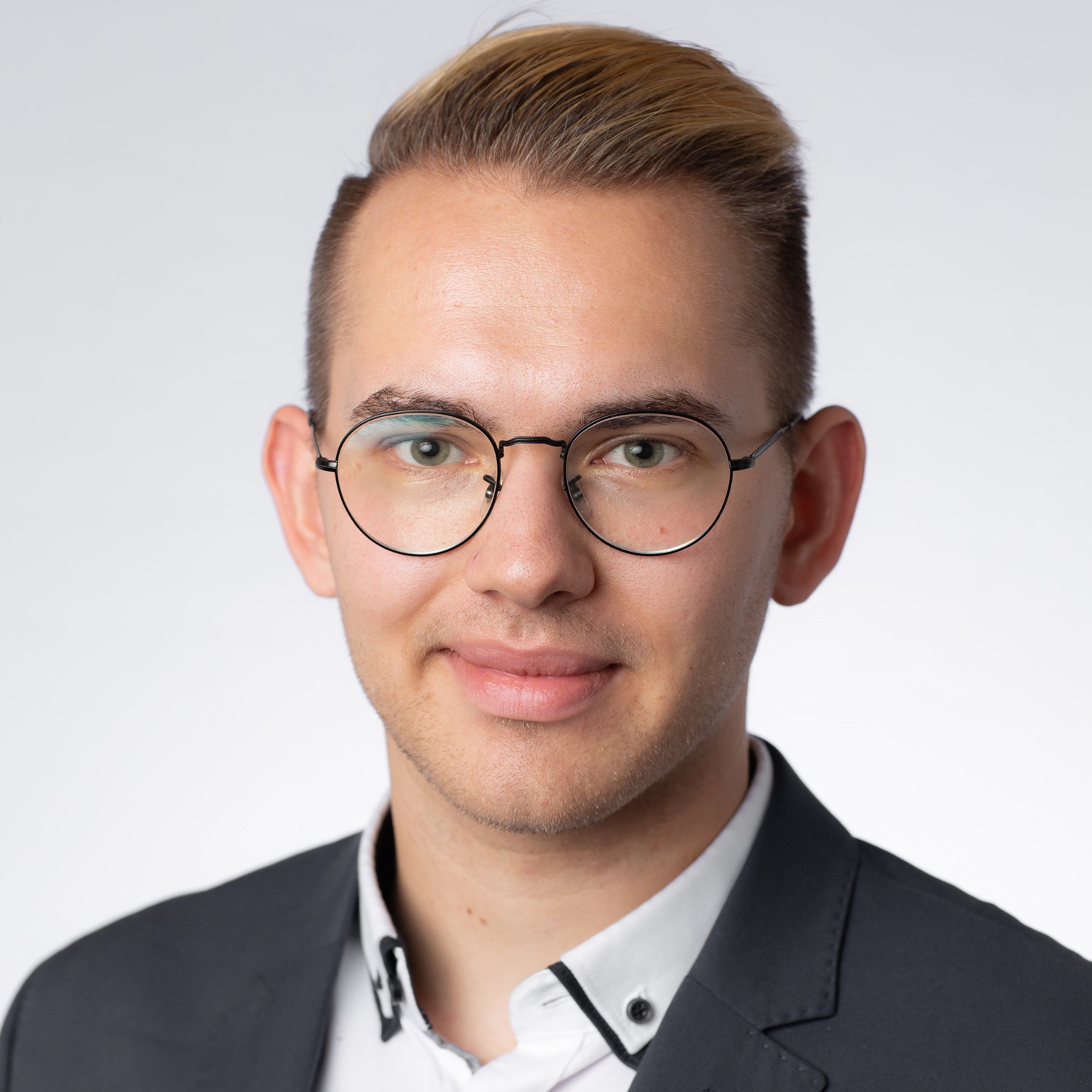}}]{Jan Kreischer} received his BSc in Informatics from Karlsruhe Institute of Technology and recently finished his master degree in Artificial Intelligence at the University of Zurich (UZH). His research interests lie in IoT, Cybersecurity, Machine Learning and Trustworthy AI.
 \end{IEEEbiography}

 % \vskip -1\baselineskip

 \begin{IEEEbiography}[{\includegraphics[width=1in,clip]{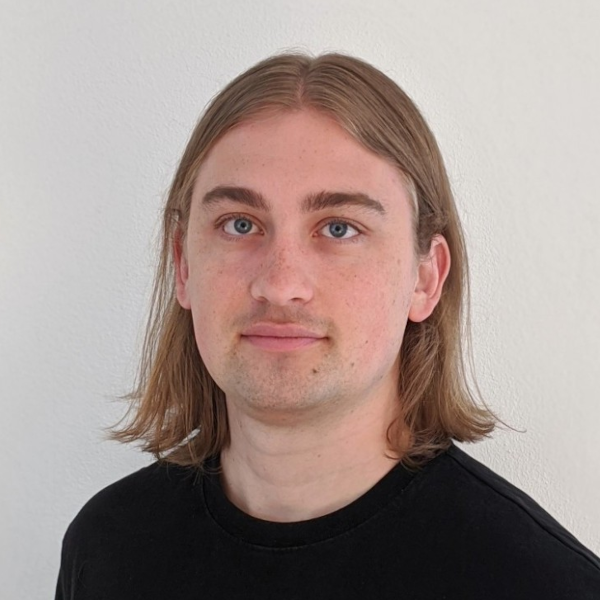}}]{Jan von der Assen} received his MSc degree in Informatics from the  University of Zurich, Switzerland. Currently, he is pursuing his Doctoral Degree under the supervision of Prof. Dr. Burkhard Stiller at the  Communication Systems Group, University of Zurich. His research interest lies at the intersection between risk management and the mitigation of cyber threats.
 \end{IEEEbiography}

 % \vskip -1\baselineskip

 \begin{IEEEbiography}[{\includegraphics[width=1in,clip]{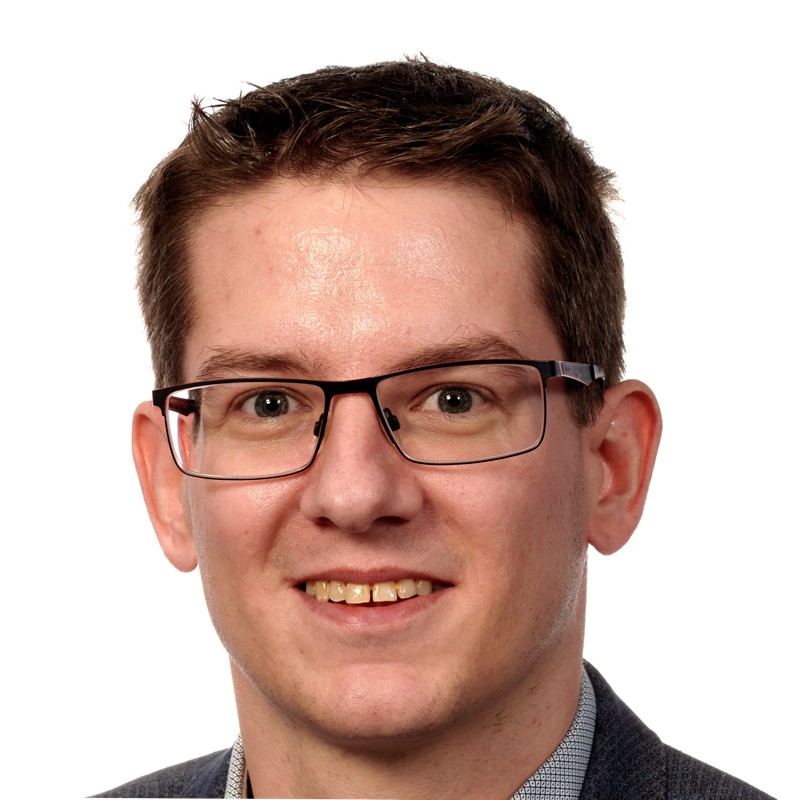}}]{Gérôme Bovet} received his Ph.D. in networks and computer systems from Telecom ParisTech, France, in 2015, and an Executive MBA from the University of Fribourg, Switzerland in 2021. He is the head of data science for the Swiss Department of Defense, where he leads a research team and portfolio of about 30 Cyber-Defence projects. His work focuses on ML and DL approaches, with an emphasis on anomaly detection, adversarial and collaborative learning applied to data gathered by IoT sensors.
 \end{IEEEbiography}
% \vspace{-90 mm}
 % \vskip -1\baselineskip

 \begin{IEEEbiography}[{\includegraphics[width=1in,clip]{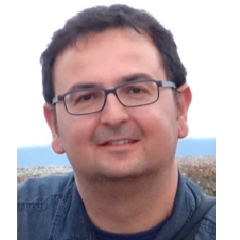}}]{Gregorio Martínez Pérez} is Full Professor in the Department of Information and Communications Engineering of the University of Murcia, Spain. His scientific activity is mainly devoted to cybersecurity and networking. He is working on different national (14 in the last decade) and European IST research projects (11 in the last decade) related to these topics, being Principal Investigator in most of them. He has published 200+ papers in international conference proceedings, magazines and journals.
 \end{IEEEbiography}
% \vspace{-90 mm}
 % \vskip -1\baselineskip

 \begin{IEEEbiography}[{\includegraphics[width=1in,clip]{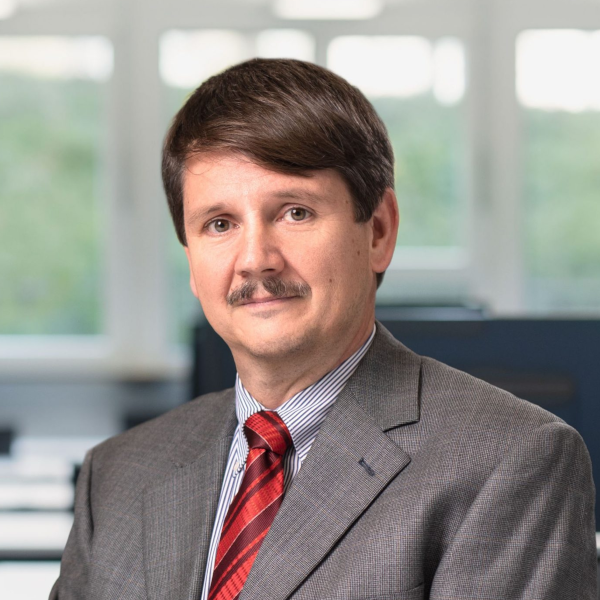}}]{Burkhard Stiller} received his MSc degree in Computer Science and the PhD degree from the University of Karlsruhe, Germany, in 1990 and 1994. Since 2004 he chairs the Communication Systems Group CSG, Department of Informatics IfI, University of Zürich UZH, Switzerland as a Full Professor. His main research interests are published in +300 papers and include decentralized systems with fully control, network and service management, IoT, and telecommunication economics.
 \end{IEEEbiography}

\end{document}